\documentclass[aps,superscriptaddress,eqsecnum,nofootinbib,preprintnumbers]{revtex4}
\usepackage{graphicx,epsfig}
\usepackage{amsmath}
\usepackage{amssymb}
\usepackage{subfigure}

\usepackage{relsize}

\newcommand{\be}{\begin{equation}}
\newcommand{\ee}{\end{equation}}
\newcommand{\ba}{\begin{eqnarray}}
\newcommand{\ea}{\end{eqnarray}}
\newcommand{\bc}{\begin{center}}
\newcommand{\ec}{\end{center}}

\newcommand{\bay}{\begin{array}{rcl}}
\newcommand{\eay}{\end{array}}

\def\h0{H_{0}}

\newcommand{\nn}{\nonumber}

\newcommand*{\LargerCdot}{\raisebox{-0.25ex}{\scalebox{1.1}{$\cdot$}}}

\begin{document}

\title{Asymptotically Safe gravity and \\
non-singular inflationary Big Bang with vacuum birth}

\author{Georgios Kofinas}
\email{gkofinas@aegean.gr} \affiliation{Research Group of Geometry,
Dynamical Systems and Cosmology\\
Department of Information and Communication Systems Engineering\\
University of the Aegean, Karlovassi 83200, Samos, Greece}

\author{Vasilios Zarikas}
\email{vzarikas@teilam.gr} \affiliation{
Central Greece University of Applied Sciences\\
Department of Electrical Engineering, 35100 Lamia, Greece}
\affiliation{Nazarbayev University, School of Engineering, Astana, Republic of Kazakhstan, 010000}

\begin{abstract}

General non-singular accelerating cosmological solutions for an initial cosmic period of pure vacuum
birth era are derived. This vacuum era is described by a varying cosmological ``constant''
suggested by the Renormalisation Group flow of Asymptotic Safety scenario near the ultraviolet fixed
point. In this scenario, natural exit from inflation to the standard decelerating cosmology occurs
when the energy scale lowers and the cosmological ``constant'' becomes insignificant.
In the following period where matter is also present, cosmological solutions with characteristics
similar to the vacuum case are generated. Remarkably the set of equations allow for particle production
and entropy generation. Alternatively, in the case of non-zero bulk viscosity, entropy production
and reheating is found. As for the equations of motion, they modify Einstein equations by adding
covariant kinetic terms of the cosmological ``constant'' which respect the Bianchi identities.
An advance of the proposed framework is that it ensures a consistent description of both a quantum
vacuum birth of the universe and a subsequent cosmic era in the presence of matter.

\end{abstract}

\maketitle

\section{Introduction}
\label{Introduction}

An interesting framework for the discovery of a theory of everything are the renormalisation group
approaches to quantum gravity \cite{perimeter} that encapsulate perturbative and non-perturbative
field theoretic techniques and functional renormalisation group flow investigations. A concrete and
minimal scheme for quantum gravity that includes no inconsistencies is the Asymptotic Safety (AS)
programm, or otherwise called Quantum Einstein Gravity \cite{reviews}. It is a model that keeps the
same fields and symmetries from General Relativity and it was first proposed as an idea by Weinberg
\cite{wein}. The key issue is the existence of a non-Gaussian fixed point (NGFP) of the renormalization
group (RG) flow for gravity. Due to this NGFP that determines the behavior of the theory
at the UV, all measured quantities are free from nonphysical divergences.

The Asymptotic Safety scenario is based on the mathematical technique of the functional
renormalization group equation for gravity \cite{mr}, which enables the detailed analysis
of the gravitational RG flow at a non-perturbative level \cite{Wetterich:1992yh}, \cite{Litim}.
It was possible to prove that the scaling behavior of the dimensionful Newtons's constant  is
antiscreened at high energies \cite{rev2}, a behavior that leads to the NGFP which is necessary
for asymptotic safety. Further studies include matter and more gravitational operators in the
action \cite{2006PhRvL..97v1301C}-\cite{ms1}.
The key ingredient of the theory is the gravitational effective average action $\Gamma_k$.
This keeps only the effect of the quantum fluctuations with momenta $p^2 > k^2$,
thus $\Gamma_k$ expresses an approximate description of physics at the momentum scale
$p^2 \approx k^2$. The truncated RG flow equations leave two running couplings (with respect to energy),
the gravitational constant $G(k)$ and the (positive) cosmological constant $\Lambda(k)$. Near the
non-Gaussian UV fixed point the coupling $G$ is known to approach a zero value, while on the other
hand, the coupling $\Lambda$ goes to infinity.

The description with the help of the effective average action and the functional RG flow enables also
the development of phenomenological investigations in the context of the asymptotic safety proposal.
Various investigations of ``RG improved" black holes were first appeared in
\cite{Bonanno:1998ye}. Other
works extended the studies to the Vaidya metric \cite{Bonanno:2006eu} and to the modified Kerr
metric \cite{Reuter:2006rg}. The thermodynamic
properties of these black holes were described in \cite{Falls:2012nd}. Black hole solutions from the
inclusion of higher derivative terms
in the effective average action were presented in \cite{Cai:2010zh}. Other works analysing
black holes with RG improvements have been proposed in \cite{Becker:2012js}-\cite{Koch:2013owa}.

Of particular interest are investigations of the nature of the microscopic structure of the
Asymptotically
Safe quantum spacetime \cite{Lauscher:2005qz}-\cite{Calcagni:2013vsa}.
It seems that quantum corrections at high energies (near the nontrivial UV fixed points)
modify drastically the classical picture since  fractal dimensionality seems to appear.

An important phenomenological topic that can also test the properties and the new point of view of
AS gravity is the study of RG improved cosmologies, first appeared in \cite{Bonanno:2001xi},
and further studied in \cite{cosmofrank}-\cite{c3} (see also review \cite{Reuter:2012xf}).
Along these lines of research it was possible to propose solution of the cosmic entropy issue
\cite{Bonanno:2008xp}. Another interesting outcome is that ``RG improved''
cosmologies admit exponential or power-law inflationary solutions
\cite{2007JCAP...08..024B}.

The scope of this work is to investigate in the context of AS gravity the
evolution of the universe at high energies.
First, we work on the assumption that cosmos had a quantum vacuum birth, first speculated in \cite{troy}.
Furthermore, the consequent period in the presence of matter is analyzed. Matter is expected to appear
due to energy transfer from vacuum to matter fields as the expansion proceeds. In summary, we
hypothesize that the universe starts in a vacuum state that is characterized by an energy dependent
cosmological constant $\Lambda(k)$. Subsequently, Einstein equations include a non-zero matter energy
momentum tensor with an energy dependent Newton's constant. Both $\Lambda$ and $G$ respect the energy
dependence that is predicted in the context of AS at the NGFP. It will be shown that the matter
solutions predict particle production and entropy generation or negative viscous pressure associated
with entropy production and reheating.

It is common in AS literature, in order to improve existing solutions of Einstein equations to set
$\Lambda(k)$ and $G(k)$ as functions of energy $k$. The simple input of $\Lambda(k)$ and
$G(k)$ into the classical vacuum equations results to violation of the Bianchi identities, while this
same input into a classical solution creates a metric which is not solution of a well-defined theory.
In \cite{bianchi}, the
formalism of obtaining RG improved solutions that respect Bianchi identities is presented at the
action level, while in \cite{Kofinas:2015sna} an alternative and mathematically more solvable
approach was developed at the level of equations of motion. In \cite{Kofinas:2015sna} the formalism
includes the appropriate covariant kinetic terms that support an arbitrary source field $\Lambda(k)$
without any symmetry assumption. Here, we extend the formalism presented in \cite{Kofinas:2015sna},
beyond the vacuum case, to also include matter. The present study provides novel quantum gravity
inspired modified Einstein equations capable to describe both absence of matter cases
and configurations where vacuum and matter contributions are realized. This new scheme proposes a
consistent way able to respect Bianchi identities in both alternatives.

Consequently, an important question is how to relate the RG scale parameter $k$ to cosmological
time/proper length in order for the differential equations to make sense.
The first works \cite{Bonanno:2001xi} have chosen in cosmology the RG
scale inversely proportional to cosmological time, and subsequently, the more favorable connection with
the Hubble scale was investigated. In other works the RG scale is linked with
the fourth root of the energy density \cite{Guberina:2002wt}, the cosmological event/particle horizons
\cite{Bauer:2005rpa}, or curvature invariants like Ricci scalar
\cite{Frolov:2011ys}-\cite{Copeland:2013vva}.

The novel scheme that is encapsulated in the presented new quantum inspired equations of motion
exhibits various new interesting features. The modified Einstein equations, together with the modified
energy-momentum conservation, suggest a constraint on the allowed/compatible matter content, and either
set a constraint or not on the allowed functional dependence $k(L)$ between the energy scale $k$ and the
geometrical scale/length $L$ that is connected to the expansion of the universe.
Remarkably both alternatives result to interesting consequences. When $k(L)$ is left free,
it is possible to get entropy generation from particle production and non-singular accelerating
solutions, while when the energy dependence $k(L)$ is restricted, similar cosmologies
with bulk viscosity, entropy generation and reheating arise.
One should notice that the presented modified Einstein equation in the spirit of AS program is an
effective description of gravity near the NGFP and they do not describe the low energy cosmic
expansion.

The organization of the paper is as follows. In section \ref{vacuum} we solve at high energies
near the NGFP the consistent RG improved equations that govern the homogeneous and isotropic
universe with energy dependent cosmological constant for different choices of the energy-length scaling.
In section \ref{matter} we present the modified equations which contain both vacuum energy and
matter and are consistent with the vacuum equations. The full space of solutions is found
with either particle production or bulk viscosity.
In section \ref{inflation} a discussion of the inflation and the thermodynamics of the universe
is presented. Finally, we conclude  in section \ref{conclusions}.


\section{Vacuum cosmological solutions}
\label{vacuum}

In \cite{Kofinas:2015sna} consistent modified Einstein equations have been presented which describe how a
classical spacetime is affected/shaped in the presence of a quantum vacuum. The quantum vacuum
is modeled through a non-zero cosmological constant term which is energy dependent according to
the AS program. Several interesting spherically symmetric solutions have also been derived there,
with some of them exhibiting non-singular behaviour. In this section we solve
the same vacuum field equations but for the case of a homogenous and isotropic metric. Vacuum solutions
with the cosmological constant as the only source are of particular importance. The reason is that the
birth of our universe from a vacuum fluctuation is a favored scenario in various quantum gravity
inspired cosmological models. An extension of the vacuum field equations appears in next
sections, capable to describe consistently the cosmic evolution including both a vacuum and a matter
content, using a positive cosmological ``constant'' $\Lambda(k)$ and a gravitational Newton's ``constant''
$G(k)$ in the spirit of AS. Thus, the same set of equations will be able to describe both cosmological
eras, namely an initial quantum vacuum birth and a subsequent period with nonzero vacuum and matter
contributions.

The proposed modified vacuum Einstein equations can be seen as a general
$\Lambda$ varying model of modified gravity. It is not related to a specific
energy dependent RG running law of the coupling $\Lambda$ and that's why it can be
useful to describe all types of running laws of $\Lambda$.
Usually in the AS literature RG improved Einstein equations (or solutions thereof)
are taken to give an effective description of physics at a characteristic energy scale $k$.
The same is true for our equations. Since our vacuum equations
contain no matter, they are perfect for the theoretical description of a semiclassical
analysis of AS spacetime near the center of black holes or in the big bang transplanckian
regime. Thus, the produced vacuum cosmological solutions do not contradict with
other proposed AS inspired cosmologies that appear in the literature; they might be seen
that describe the very initial period of the universe before the period
described by other asymptotic safe gravity cosmologies that appear in the literature.

Let us begin modeling the quantum vacuum dominated initial cosmic era. The mathematical description of
this era is based on the modified vacuum Einstein equations derived in \cite{Kofinas:2015sna}
\begin{equation}
G_{\mu\nu}=-\bar{\Lambda}\,e^{\psi} g_{\mu\nu}
-\frac{1}{2}\psi_{;\mu}\psi_{;\nu}-\frac{1}{4}g_{\mu\nu}\psi^{;\rho}
\psi_{;\rho}+\psi_{;\mu;\nu}-g_{\mu\nu}\Box\psi\,.
\label{vaceinstein}
\end{equation}
The field $\psi(x)$ is related to the cosmological constant through $\Lambda=\bar{\Lambda}\,e^{\psi}$,
where $\bar{\Lambda}$ is an arbitrary constant reference value. Equations (\ref{vaceinstein}) form
a minimal extension of Einstein equations containing first and second derivatives of $\psi$.
They are by construction identically covariantly conserved for any $\psi(x)$, so the Bianchi
identities are satisfied.
Since $\psi(x)$ does not have its own equation of motion, it can be determined externally,
e.g. as implied by the AS scenario. Indeed, geometry independent RG flow equations predict the
running of both $\Lambda(k)$, $G(k)$ at high
energies near the NGFP, where the cosmological constant/coupling is given by \cite{Manrique:2010am}
\begin{equation}
\Lambda=\lambda_{\ast}k^{2}
\label{ejk}
\end{equation}
with $\lambda_{\ast}>0$ a dimensionless constant.

We will be interested in the present work in the investigation of cosmology at very high energies,
so that equation (\ref{ejk}) can be applied.
The spatially homogeneous and isotropic cosmological metric is
\begin{equation}
ds^{2}=-n(t)^{2}dt^{2}+a(t)^{2}\Big[\frac{dr^{2}}{1\!-\!\kappa\,r^{2}}+r^{2}\big(d\theta^{2}
\!+\!\sin^{2}{\!\theta}\,d\phi^{2}\big)\Big]\,,
\label{jkw}
\end{equation}
where $n(t)$ is the lapse function and $\kappa=-1,0,1$ characterizes the spatial curvature.
The coupling $\Lambda(x)$ carries the same symmetries, so it is $\Lambda=\Lambda(t)$. For the metric
(\ref{jkw}) the non-vanishing components of the Einstein tensor $G_{\nu}^{\mu}$ are
\begin{eqnarray}
G^{t}_{t}&=&-3\Big(H^{2}+\frac{\kappa}{a^{2}}\Big) \nn \\
G^{i}_{j}&=&-\Big(\frac{2}{n}\dot{H}+3H^{2}+\frac{\kappa}{a^{2}}\Big)\delta^{i}_{j},
\label{skk}
\end{eqnarray}
where the indices $i,j$ refer to the spatial coordinates, $H=\frac{\dot{a}}{na}$ is the Hubble
parameter and a dot denotes a differentiation with respect
to $t$. It is also possible to evaluate
\begin{equation}
\label{psi1}
\psi^{;\rho}\psi_{;\rho}=-\frac{\dot{\psi}^{2}}{n^{2}}\,\,\,\,,\,\,\,\,
\Box\psi=-\frac{1}{n}\Big(\frac{\dot{\psi}}{n}\Big)^{^{\LargerCdot}}-3H\frac{\dot{\psi}}{n}
\end{equation}
and the non-vanishing components of $\psi_{;\mu;\nu}$ by
\begin{equation}
\label{comps}
\psi_{;t;t}=n\Big(\frac{\dot{\psi}}{n}\Big)^{^{\LargerCdot}}\,\,\,\,,\,\,\,\,
\psi_{;r;r}=-\frac{Ha^{2}}{1\!-\!\kappa\,r^{2}}\,\frac{\dot{\psi}}{n}\,\,\,\,,\,\,\,\,
\psi_{;\theta;\theta}=-H a^{2}r^{2}\frac{\dot{\psi}}{n}\,\,\,\,,\,\,\,\,
\psi_{;\phi;\phi}=-H a^{2}r^{2}\sin^{2}\!\theta\,\frac{\dot{\psi}}{n}\,.
\end{equation}
Therefore, the two components of (\ref{vaceinstein}) are
\begin{eqnarray}
3\Big(H^{2}+\frac{\kappa}{a^{2}}\Big)
=\bar{\Lambda}e^{\psi}-3H\frac{\dot{\psi}}{n}-\frac{3\dot{\psi}^{2}}{4n^{2}}
\label{eqvac1}\\
\frac{2}{n}\dot{H}+3H^{2}+\frac{\kappa}{a^{2}}
=\bar{\Lambda}e^{\psi}-2H\frac{\dot{\psi}}{n}-\frac{\dot{\psi}^{2}}{4n^{2}}
-\frac{1}{n}\Big(\frac{\dot{\psi}}{n}\Big)^{^{\LargerCdot}}.
\label{eqvac2}
\end{eqnarray}
Equations (\ref{eqvac1}), (\ref{eqvac2}) are satisfied by construction for any $\psi(t)$.
Equation (\ref{eqvac2}) is not independent, since by differentiating equation (\ref{eqvac1}) with
respect to $t$ and using (\ref{eqvac1}) itself, we get (\ref{eqvac2}) multiplied by
$1\!+\!\frac{\dot{\psi}}{2nH}$. Hereafter, in the following vacuum solutions we consider $t$
to be the cosmic time and take $n=1$.
In order to proceed with the solution of (\ref{eqvac1}) we have to
determine $\psi$ as a function of e.g. $t,H,a$, using equation (\ref{ejk}).
This will come by the selection of a scaling that associates the
energy of RG scale $k$ to a characteristic time or length of the solution.

It worths to mention that $k$ as a function of cosmic time starts with an infinite or a very high
value and should decrease as the universe departs from the NGFP. However, it is possible
that during the cosmic evolution, still in the proximity of the NGFP, $k(t)$ may increase for
some era.

\subsection{Scaling $k\propto 1/t$}
\label{inverse time}

In cosmological models of Asymptotically Safe gravity it is common to use as a reasonable scaling
the following expression \cite{Bonanno:2001xi}, \cite{cosmofrank}
\begin{equation}
k=\frac{\xi}{t}\,,
\label{scale1a}
\end{equation}
where $\xi>0$ is a dimensionless parameter and time $t$ is considered positive valued.
We are interested in understanding the cosmological behaviour in a regime where time takes
sufficiently small values so that $k$ takes its high energy values.
From (\ref{scale1a}) obviously $k$ decreases with time.
The relation of $\psi$ with time is
$e^{\psi}=\lambda_{\ast}\xi^{2}/(\bar{\Lambda}t^{2})$
and the Friedmann equation (\ref{eqvac1}) becomes
\begin{equation}
H^{2}+\frac{\kappa}{a^{2}}=\frac{\lambda_{\ast}\xi^{2}\!-\!3}{3t^{2}}+\frac{2H}{t}\,.
\label{gye}
\end{equation}
Equation (\ref{gye}) is invariant under the transformation $t\rightarrow \lambda t$,
$a\rightarrow \lambda a$, therefore defining
\begin{equation}
z=\frac{a}{t}\,,
\label{ner}
\end{equation}
we find the equation
\begin{equation}
\dot{z}^{2}=\frac{\omega^{2}z^{2}-\kappa}{t^{2}}\,,
\label{yns}
\end{equation}
where
\begin{equation}
\omega=\sqrt{\frac{\lambda_{\ast}\xi^{2}}{3}}\,.
\label{gyw}
\end{equation}

For $\kappa=0$, the solution is
\begin{equation}
a(t)=ct^{1\pm\omega}\,,
\label{hgr}
\end{equation}
where $c>0$ is integration constant.
For the upper branch, or for the lower branch with $\omega<1$, the solutions are expanding
starting for $t=0$ at zero scale factor. These solutions have divergent Ricci scalar $R$ at $t=0$,
therefore they are typical singular solutions.
Note that the upper branch describes a power law inflation which becomes stronger as $\omega\gg 1$.
Especially the lower branch for $\omega=\frac{1}{2}$ has $R=0$ for $t=0$,
but the divergencies appear in higher curvature invariants.
For the lower branch with $\omega>1$ the solution is contracting starting for $t=0$ at infinite
scale factor.
An interesting general comment can be made at this point. In the AS scenario, an inflationary period
is always followed by a natural exit from inflation and this occurs when
the energy scale becomes smaller than the inflation scale.
Then, the quantum modifications of $\Lambda(k)$ start to become insignificant and
its value is rapidly decreasing, leading to the standard decelerating cosmology.

For $\kappa=+1$ it should be from (\ref{yns}) $a>\omega^{-1}t$.
The solution of (\ref{yns}) gives
\begin{equation}
a(t)=\frac{t}{2\omega}\big(ct^{\pm\omega}+c^{-1}t^{\mp\omega}\big)\,,
\label{tnd}
\end{equation}
where $c>0$ is integration constant, under the constraint $a<c\omega^{-1}t^{1\pm\omega}$, i.e.
$ct^{\pm\omega}>1$.
Due to this constraint it is seen that the upper branch describes a non-singular (finite
curvature invariants)
expanding universe with $a_{\text{min}}=\omega^{-1}c^{-\frac{1}{\omega}}$,
$t_{\text{min}}=c^{-\frac{1}{\omega}}$,
where the maximum value $k_{\text{max}}=\xi c^{\frac{1}{\omega}}$ can be
made as large as desired choosing $c$ sufficiently large.
Moreover, it is seen that this solution has $\ddot{a}>0$ which means that it is accelerating.
For the lower branch with $\omega<1$, the solution is expanding
starting for $t=0$ at zero scale factor; at $t=0$ there is a singularity with divergent Ricci
scalar and close to $t=0$ it is to leading order $a\approx \frac{c}{2\omega}t^{1-\omega}$.
Moreover, this universe enters from a decelerating to an accelerating phase.
The lower branch for $\omega>1$ describes a universe which starts from infinite volume, collapses,
and at a finite scale factor bounces to an expanding universe which has an end; at the bounce it is
$t\sim c^{\frac{1}{\omega}}$, $a\sim c^{\frac{1}{\omega}}$ and the energy scale
$k\sim \xi c^{-\frac{1}{\omega}}$ can be as large as desired choosing $c$ sufficiently small.
This branch is also accelerating.
The lower branch with $\omega=1$ describes also a non-singular expanding universe.

Finally, for $\kappa=-1$, the solution is
\begin{equation}
a(t)=\frac{t}{2\omega}\big(ct^{\pm\omega}-c^{-1}t^{\mp\omega}\big)\,,
\label{hgg}
\end{equation}
where $c>0$ is integration constant.
Since $a>0$ it should be $ct^{\pm\omega}>1$. For the upper branch the solution is expanding starting
for $t=c^{-\frac{1}{\omega}}$ at zero scale factor where there is a curvature singularity; this
solution is accelerating. For the lower branch with $\omega<1$ the solution
initially expands from a singularity starting for $t=0$ at zero scale factor and finally
bounces and collapses again to a singular zero volume; this solution is decelerating.
For the lower branch with $\omega>1$ the solution is contracting starting for
$t=0$ at infinite scale factor and results to a singular big crunch.

We resume with the most interesting general solutions of the scaling $k\propto 1/t$.
For the spatially flat 3-space topology a strong power law inflation can occur close to the
initial singularity. For the positively curved case all solutions have accelerating
phases which can support an inflationary epoch;
they either avoid the initial big bang singularity or they possess a big bang or during
a collapsing phase avoid the big crunch towards expansion. For the negatively curved topology
a singular accelerating cosmology can appear.

For the alternative scaling $k=\frac{\xi}{a(t)}$, where $k$ is inversely proportional to the proper
distance at fixed $t$, equation (\ref{eqvac1}) is satisfied for any $a(t)$ given that
$\kappa=1=\frac{\lambda_{\ast}\xi^{2}}{3}$, so it does not provide useful information.
This is also an outcome of other AS cosmology studies found in literature
\cite{Bonanno:2001xi}, \cite{cosmofrank}.

\subsection{Scaling $k\propto H(t)$}
\label{hubble scale}

In order to investigate the time evolution of the cosmic scale factor using the full effective
action $\Gamma(g_{\mu\nu})$, it can be made use of the fact that the Hubble parameter appears as a
mass in propagators.
Therefore, a sensible approximation is to disregard the contributions of quantum fluctuations with
wavelengths greater than $H^{-1}$ since they are suppressed. This leads to use as a
connection of energy scale to the length scale a relation of the form $k\sim H(t)$ \cite{cosmofrank}.
Thus, we assume here
\begin{equation}
k=\xi H(t)\,,
\label{scale1}
\end{equation}
where the dimensionless parameter $\xi$ is $\xi>0$ for $H>0$ and $\xi<0$ for $H<0$.
It is obvious that a bouncing solution is not possible in this case.
Then, $e^{\psi}=\lambda_{\ast}\xi^{2}/(\bar{\Lambda}H^{2})$
and the Friedmann equation (\ref{eqvac1}) becomes
\begin{equation}
(1\!-\!\omega^{2})H^{2}+2\dot{H}+\frac{\dot{H}^{2}}{H^{2}}+\frac{\kappa}{a^{2}}=0\,,
\label{hyd}
\end{equation}
where
\begin{equation}
\omega=\sqrt{\frac{\lambda_{\ast}\xi^{2}}{3}}\,.
\label{hue}
\end{equation}
Equation (\ref{hyd}) is written as
\begin{equation}
a^{2}\ddot{a}^{2}-\omega^{2}\dot{a}^{4}+\kappa\dot{a}^{2}=0\,.
\label{heu}
\end{equation}
Setting
\begin{equation}
u=\dot{a}\,,
\label{deg}
\end{equation}
equation (\ref{heu}) takes the form
\begin{equation}
a\frac{du}{da}=\pm\sqrt{\omega^{2}u^{2}\!-\!\kappa}\,.
\label{ijm}
\end{equation}
If $\kappa=1$ it should be $|u|>\omega^{-1}$.

For $\kappa=0$ the solution of (\ref{ijm}) is
\begin{equation}
u=ca^{\pm\omega}\,,
\label{tne}
\end{equation}
with $c$ an integration constant, and thus
\begin{equation}
a(t)=\big[c(1\!\mp\!\omega)(t\!-\!t_{0})\big]^{\frac{1}{1\mp\omega}}\,,
\label{ndt}
\end{equation}
where $t_{0}$ is an integration constant. The constant $c$ should be positive for expanding solutions.
The upper branch with $\omega<1$ or the lower branch describe typical power law singular expanding
solutions. However, this upper branch is inflationary and the inflation can become very strong if
$\omega \rightarrow 1$. Moreover, in both cases it can be seen that $k$ decreases with time.
The upper branch with $\omega>1$ describes a collapsing universe which asymptotically goes to
zero scale factor with finite however curvature invariants.
\newline
Especially for the upper branch with $\omega=1$ we get the non-singular DeSitter solution
$a\propto e^{ct}$ which describes a typical inflationary period with the main advance, as referred
previously, that a natural exit occurs.

For $\kappa=+1$ the solution of (\ref{ijm}) is
\begin{equation}
u=\frac{1}{2\omega}\big(ca^{\pm\omega}+c^{-1}a^{\mp\omega}\big)\,,
\label{ney}
\end{equation}
where $c>0$ is integration constant.
Since $u>0$, there are only expanding solutions and it is $\xi>0$.
For the upper branch it is $a>c^{-\frac{1}{\omega}}$ which
means that the solution avoids the zero scale factor regime.
It is obvious from equations (\ref{scale1}), (\ref{deg}), (\ref{ney})
that this solution has a finite $k_{\text{max}}$.
Moreover, it can be shown that the energy scale $k$ decreases with time.
For the lower branch it is $a<c^{\frac{1}{\omega}}$.
\newline
Using (\ref{ney}) it can be found the Ricci scalar as a function of the scale factor to be
\begin{equation}
R=\frac{3}{2\omega^{2}a^{2}}\big[c^{2}(1\!\pm\!\omega)a^{\pm 2\omega}+c^{-2}(1\!\mp\!\omega)
a^{\mp 2\omega}+2(1\!+\!2\omega^{2})\big]\,.
\label{eor}
\end{equation}
Therefore, the upper branch has finite scalar curvature and leads to a non-singular cosmology,
while the lower branch has a curvature singularity at $a=0$.
From (\ref{ijm}) it is seen that the upper branch is accelerating and the lower decelerating.
\newline
Integrating (\ref{ney})
{\footnote{The differential equation (\ref{ney}) or (\ref{ney1}) $2\omega\dot{a}=ca^{\pm\omega}
+\kappa c^{-1}a^{\mp\omega}$ is integrated through the transformation $x=c^{2}a^{\pm 2\omega}$
to $t-t_{0}=\pm c^{\mp\frac{1}{\omega}}\int\frac{x^{\alpha}}{x+\kappa}dx$,
thus $t-t_{0}=\pm c^{\mp\frac{1}{\omega}}\frac{x^{\alpha+1}}
{\alpha+1}f(-\kappa x)$, where $f(x)$ satisfies equation (\ref{jie}).}},
the dependence on time can be obtained as
\begin{equation}
t-t_{0}=\frac{2c\omega}{1\pm\omega} a^{1\pm\omega} f(-c^{2}a^{\pm 2\omega})\,,
\label{ksr}
\end{equation}
where $t_{0}$ is an integration constant which can be absorbed into a redefinition of $t$.
Here the function $f(x)$ satisfies the hypergeometric differential equation
\begin{equation}
x(1\!-\!x)\frac{d^{2}f(x)}{dx^{2}}
+\big[\alpha\!+\!2-(\alpha\!+\!3)x\big]\frac{df(x)}{dx}-(\alpha\!+\!1)f(x)=0\,,
\label{jie}
\end{equation}
where $\alpha=\frac{\pm 1-\omega}{2\omega}$.
Equation (\ref{jie}) has (for $\alpha$ not an integer) two independent solutions
\footnote{see \cite{erdelyi}, p.71, formula 8.},
one is $x^{-\alpha-1}$ and the
other $_{2}F_{1}(1,\alpha+1;\alpha+2;x)$. The first solution just contributes to the constant
$t_{0}$, thus
\begin{equation}
t-t_{0}=\sigma\frac{2c\omega}{1\pm\omega} a^{1\pm\omega} \, _2F_1\Big(1,\frac{\omega\pm 1}{2\omega};
\frac{3\omega\pm 1}{2\omega};-c^{2}a^{\pm 2\omega}\Big)\,,
\label{drm}
\end{equation}
where $\sigma$ is a proportionality constant to be determined from some limiting process where the
time integral can be computed exactly and it arises due to that the hypergeometric equation is
homogeneous.
\newline
For the upper branch, since $a>c^{-\frac{1}{\omega}}$, in the limit $a\rightarrow \infty$ it can be
found the behaviour $t-t_{0}'\approx \frac{2c^{-1}\omega}{1-\omega}a^{1-\omega}$. On the other hand,
the hypergeometric function in (\ref{drm}) is expressed \footnote{see \cite{abramowich},
p.559, formula 15.3.7.} as $_{2}F_{1}(1,\alpha+1;\alpha+2;-x)=\frac{\alpha+1}{\alpha}\frac{1}{x}
\,_{2}F_{1}(1,-\alpha;1-\alpha;-\frac{1}{x})+\frac{\Gamma(\alpha+2)\Gamma(-\alpha)}
{x^{\alpha+1}}$, therefore for $x\rightarrow +\infty$ it is $_{2}F_{1}(1,\alpha+1;\alpha+2;-x)
\approx \frac{\alpha+1}{\alpha}\frac{1}{x}+\frac{\Gamma(\alpha+2)\Gamma(-\alpha)}{x^{\alpha+1}}$.
Comparing the two asymptotic expressions it is found $\sigma=1$ and (\ref{drm}) becomes
\begin{equation}
t-t_{0}=\frac{2c\omega}{1+\omega} a^{1+\omega} \, _2F_1\Big(1,\frac{\omega+1}{2\omega};
\frac{3\omega+1}{2\omega};-c^{2}a^{2\omega}\Big)\,.
\label{drd}
\end{equation}
For $\omega<1$ this solution expresses a non-singular universe starting at a finite $t$ and
expanding to infinity at infinite
proper time, while for $\omega>1$ it is again a non-singular universe expanding to infinity
in finite proper time, so it develops a big rip (of course this big rip is not true since
the validity of the approximation terminates after some time).
\newline
For the lower branch, since $a<c^{\frac{1}{\omega}}$, in the limit $a\rightarrow 0$ it can be found
the behaviour $t-t_{0}'\approx \frac{2c^{-1}\omega}{1+\omega}a^{1+\omega}$. Similarly as before,
approximating the hypergeometric function in the neighborhood of $a=0$, we find $\sigma=1$, thus
(\ref{drm}) becomes
\begin{equation}
t-t_{0}=\frac{2c\omega}{1-\omega} a^{1-\omega} \, _2F_1\Big(1,\frac{\omega-1}{2\omega};
\frac{3\omega-1}{2\omega};-c^{2}a^{-2\omega}\Big)\,.
\label{rir}
\end{equation}
This solution represents a singular expanding cosmology.
\newline
For $\omega=1$ the above analysis does not work since $\alpha$ becomes integer. In this case the
analytic solution of (\ref{ney}) is simpler, $a=c^{\mp 1}\sqrt{e^{c^{\pm 1}(t-t_{0})}-1}$. The upper
branch is an expanding non-singular solution and the lower an expanding singular one.

For $\kappa=-1$ the solution of (\ref{ijm}) is
\begin{equation}
u=\frac{1}{2\omega}\big(ca^{\pm\omega}-c^{-1}a^{\mp\omega}\big)\,,
\label{ney1}
\end{equation}
where $c>0$ is integration constant. In this case there is no bound for $a$.
Integrating (\ref{ney1}) the dependence on time can be obtained as
\begin{equation}
t-t_{0}=\frac{2c\omega}{1\pm\omega} a^{1\pm\omega} f(c^{2}a^{\pm 2\omega})\,,
\label{ksb}
\end{equation}
where $f(x)$ satisfies again the differential equation (\ref{jie}).
Similarly to the case $\kappa=1$ it arises for $\alpha$ non-integer
\begin{equation}
t-t_{0}=\sigma\frac{2c\omega}{1\pm\omega}a^{1\pm\omega}\, _2F_1\Big(1,\frac{\omega\pm 1}{2\omega};
\frac{3\omega\pm 1}{2\omega};c^{2}a^{\pm2 \omega}\Big)\,.
\label{dth}
\end{equation}
There are no expanding solutions in this case, since for $H>0$ it is from (\ref{ney1}) that
$ca^{\pm\omega}>1$, but then, the argument of the hypergeometric function in (\ref{dth}) is
larger than 1 and the hypergeometric function gets complex values. Concerning the collapsing
solutions, for the upper branch it is $a<c^{-\frac{1}{\omega}}$ and from the limiting behaviour
for $a\approx 0$ it arises $\sigma=-1$, thus the collapsing solution starts with finite initial
scale factor with $t\rightarrow -\infty$ and results to $a=0$ at finite time.
Concerning the collapsing solutions of the lower branch it is $a>c^{\frac{1}{\omega}}$
and from the limiting behaviour for $a\rightarrow\infty$ it arises $\sigma=-1$. Thus for
$\omega>1$ the collapsing solution starts with infinite initial scale factor at finite time
and results to finite scale factor at infinite $t$. For $\omega<1$ the solution starts with an
infinite scale factor at $t=-\infty$ and results to finite $a$ with $t=+\infty$.

We summarize with the most interesting solutions of the scaling $k\propto H$.
For the spatially flat 3-space topology a strong power law inflation can happen close to the
initial singularity. For the positively curved case there are general expanding solutions which
are non-singular and accelerating, so they can support inflation.

\section{Cosmological matter solutions}
\label{matter}

In this section we add, beyond the spacetime-dependent cosmological constant $\Lambda(x)$, some
extra matter carrying an energy-momentum tensor $T_{\mu\nu}$. The full gravitational equation is now
obtained by supplementing the vacuum equations (\ref{vaceinstein}) with $T_{\mu\nu}$ in the
following minimal way
\begin{equation}
G_{\mu\nu}=-\bar{\Lambda}\,e^{\psi} g_{\mu\nu}
-\frac{1}{2}\psi_{;\mu}\psi_{;\nu}-\frac{1}{4}g_{\mu\nu}\psi^{;\rho}
\psi_{;\rho}+\psi_{;\mu;\nu}-g_{\mu\nu}\Box\psi+8\pi G T_{\mu\nu}\,.
\label{skl}
\end{equation}
The Newton's constant $G$ will also be assumed to be spacetime-dependent, $G(x)$, while again the field
$\psi(x)$ is defined through $\Lambda=\bar{\Lambda}\,e^{\psi}$ with $\bar{\Lambda}$ an arbitrary
constant reference value. In the scenario of
asymptotic safety both $\Lambda,G$ are supposed to be determined uniquely as functions of the
energy scale from the RG flow equations, so it is $\Lambda(k), G(k)$, but at present we construct a
formulation where $\Lambda,G$ can be kept arbitrary functions. Later, we will employ for the early-times
cosmological period the energy dependent couplings predicted at the NGFP of AS scenario to be
\begin{equation}
\Lambda=\lambda_{\ast}k^{2}\,\,\,\,\,\,,\,\,\,\,\,\, G=\frac{g_{\ast}}{k^{2}}\,,
\label{scaling}
\end{equation}
where $\lambda_{\ast},g_{\ast}>0$ are dimensionless constants (the above scalings are also
consistent with dimensional analysis without the introduction of a new energy scale). Since in the
absence of $T_{\mu\nu}$, equation (\ref{skl}) is identically covariantly conserved for any $\psi(x)$,
the following conservation equation for $T_{\mu\nu}$ holds
\begin{equation}
(GT_{\mu\nu})^{;\nu}=0\,,
\label{kem}
\end{equation}
which provides an interaction between $T_{\mu\nu}$ and $G$. This way, equation (\ref{skl}) is
meaningful either in the absence or in the presence of a matter content.
The main advance of our theory is indeed that it encapsulates the vacuum case, something that has not
appeared so far in the literature of varying constants or AS gravities.

An alternative option, assumed often in the literature, other than equation (\ref{kem}), would be
to ignore the $\psi$-kinetic terms in (\ref{skl}), and then, the Bianchi identities would imply another
conservation equation for $T_{\mu\nu}$ containing both $\Lambda,G$ and their derivatives.
Another option, also met in the literature, would be, besides ignoring again the $\psi$-kinetic
terms, to assume the exact conservation of matter $T_{\mu\nu}^{\,\,\,\,\,\,;\nu}=0$, and then,
$\Lambda,G$ cannot be picked arbitrarily and are usually incompatible with AS relation
$G\sim \Lambda^{-1}$. However, both these two approaches are not satisfactory since in the absence
of matter the system would be inconsistent.
Considering a variation of an action principle, equations containing different than (\ref{skl})
$\psi$-kinetic terms, as well as $G$-kinetic terms, could in principle arise, however, such
a scheme cannot be explicitly implemented for a general case matter content $T_{\mu\nu}$.
Finally, in the literature \cite{Hindmarsh:2011hx}, \cite{Ahn:2011qt} another approach considers
specific form for the metric (cosmological, spherically symmetric, e.tc.),
solves some RG-like differential equations together with the Einstein equations and
finds different than (\ref{scaling}) functional dependences $\Lambda(k),G(k)$.
In the scheme that we follow we have the advantage that we use the functions (\ref{scaling})
suggested by the background independent \cite{mr} running of RG equations.

As in the previous section with the vacuum solutions, we also here assume a spatially homogeneous
and isotropic metric for the cosmic spacetime of the form (\ref{jkw}). Since the external fields
$\Lambda(x),G(x)$ carry the same symmetries, they will be of the form $\Lambda(t),G(t)$.
We consider a diagonal energy-momentum tensor $T^{\mu}_{\nu}$, so we take as matter
content a non-perfect fluid with energy density $\rho$, thermodynamic pressure $p$ and a
non-equilibrium part $\pi$ \cite{Weinbergold}-\cite{tetrdis}.
Due to the working symmetries of isotropy and homogeneity shear viscosity and energy fluxes
are disregarded.
The energy momentum tensor is
\begin{equation}
T^{\mu\nu}=\rho u^{\mu}u^{\nu}+(p+\pi)(g^{\mu\nu}\!+\!u^{\mu}u^{\nu})
\label{tmn}
\end{equation}
with $u^{\mu}$ the fluid 4-velocity. The extra pressure
$\pi$ can either be associated to a pressure due to particle production/destruction or to a bulk
viscous pressure. In the next two subsections these two cases will be discussed together with the
assumption (\ref{scaling}) of AS at the very early high energy universe. The term $\pi$ could indeed be
important during the transition phase that connects the quantum vacuum stage of the universe to the
subsequent era with non-zero matter density.

The two independent components of (\ref{skl}) are
\begin{eqnarray}
3\Big(H^{2}+\frac{\kappa}{a^{2}}\Big)
=\bar{\Lambda}e^{\psi}-3H\frac{\dot{\psi}}{n}-\frac{3\dot{\psi}^{2}}{4n^{2}}
+8\pi G\rho
\label{wkd}\\
\frac{2}{n}\dot{H}+3H^{2}+\frac{\kappa}{a^{2}}
=\bar{\Lambda}e^{\psi}-2H\frac{\dot{\psi}}{n}-\frac{\dot{\psi}^{2}}{4n^{2}}
-\frac{1}{n}\Big(\frac{\dot{\psi}}{n}\Big)^{^{\LargerCdot}}-8\pi G P\,,
\label{sdf}
\end{eqnarray}
where the total effective pressure is $P=p+\pi$.
The conservation equation (\ref{kem}) takes the form
\begin{equation}
\dot{\rho}+3nH(\rho+P)+\rho\frac{\dot{G}}{G}=0\,.
\label{elk}
\end{equation}

The system of equations (\ref{wkd})-(\ref{sdf}) is satisfied by construction for any $\psi(t)$.
However, the effective pressure $P$ is constrained to be $P=-\frac{1}{3}\rho$.
Indeed, differentiating (\ref{wkd}) with respect to $t$ and
using (\ref{elk}) to substitute $\dot{\rho}$ and also (\ref{wkd}) itself we find
\begin{equation}
\Big(1\!+\!\frac{\dot{\psi}}{2nH}\Big)
\Big[\frac{2}{n}\dot{H}\!+\!3H^{2}\!+\!\frac{\kappa}{a^{2}}
\!-\!\bar{\Lambda}e^{\psi}\!+\!2H\frac{\dot{\psi}}{n}\!+\!\frac{\dot{\psi}^{2}}{4n^{2}}
\!+\!\frac{1}{n}\Big(\frac{\dot{\psi}}{n}\Big)^{^{\LargerCdot}}\!+\!8\pi GP\Big]
=4\pi G\frac{\dot{\psi}}{nH} \Big(P\!+\!\frac{1}{3}\rho\Big)\,.
\label{lej}
\end{equation}
From (\ref{sdf}), (\ref{lej}) it arises the constraint
\begin{equation}
p+\frac{1}{3}\rho+\pi=0\,.
\label{constr}
\end{equation}
Therefore, given this consistency condition, equation (\ref{sdf}) is redundant due to the time
reparametrization and can be omitted. So, we remain with equations (\ref{wkd}), (\ref{elk}),
(\ref{constr}) and the gauge $n=1$ will be adopted.

We find that the constraint (\ref{constr}) is an interesting issue of the proposed framework,
providing restrictions on the acceptable forms of equations of state. Although it may look
as a disadvantage at first sight, however such constraints could be desired in modified
gravities and orient us to the specification of the physical content of the theory.
It is well known that in the context of conventional quantum field theory, renormalizability poses
constraints in the allowed fields or interactions. Physics is still unknown in the transplanckian
regime where our solutions are supposed to be valid and it may be that this constraint equation provides
an insight about this regime through handable equations. There may also be a concern how one will be able, following our framework,
to retrieve classical FRW cosmology, since equation (\ref{constr}) restricts some equations of states
of matter that appear afterwards. The answer to this puzzle is that the proposed
framework is valid only near the NGFP and not afterwards at lower energies.
It is expected that the transition from the transplanckian regime to the classical regime will
be explained through a quantum process of the type of decoherence on the ensemble of
superimposed quantum spacetimes instead of a set of modified Einstein equations.
Thus, such a transition requires to be modeled with a set
of quantum field equations of motion. Anyway, asymptotic safety scenario guarantees that
later on we recover pure GR equations without modification. Our equations do not describe
neither the passage regime nor the FRW period.
We note that one should always keep in mind that at large $k$ the
existence of a NGFP is more a mathematical issue, as it provides a starting point
for a well-defined quantum-gravitational path integral. The fundamental Lagrangian and the
underlying physical processes are not yet known. It is expected that a quantum gravity
interpretation will be fully resolved/understood when it will be possible to analyse how
the classical spacetime at low energies arises from the quantum ensemble of $g_{\mu\nu}$
states in a meaningful way, allowing a measurement theory in the fully quantum regime where
no classical time and clocks exist. The present work proposes a new framework to obtain
modified Einstein equations with varying $\Lambda,G$ and finds non-singular spacetimes
assuming that $\Lambda$, $G$ have the behaviour proposed by AS near the NGFP; such classical
non-singular solutions should contribute only at the quantum-gravitational path integral.

Note that for simplicity we have not included $G$-kinetic terms in (\ref{skl}). In a more
complete treatment, however, such terms should be added.
If we parametrize $G$ by $G=\bar{G}\,e^{\chi}$, an extra energy-momentum tensor $\vartheta^{(G)}_{\mu\nu}$
of $G$ would be added on the right hand side of (\ref{skl}), constructed out of $\chi$ and its first
and second derivatives
\begin{equation}
\vartheta^{(G)}_{\mu\nu}=A(\chi)\chi_{;\mu}\chi_{;\nu}+B(\chi)g_{\mu\nu}\chi^{;\rho}
\chi_{;\rho}+C(\chi)\chi_{;\mu;\nu}+E(\chi)g_{\mu\nu}\Box\chi+F(\chi)g_{\mu\nu}\,.
\label{heir}
\end{equation}
The Bianchi identities imply $\theta_{\mu\nu}^{(G);\mu}=0$, which provides the following
equations, following the process appeared in \cite{Kofinas:2015sna}
\begin{equation}
A=C'-\frac{1}{2}C^{2}\,\,\,\,\,,\,\,\,\,\,B=-C'-\frac{1}{4}C^{2}\,\,\,\,\,,\,\,\,\,\,E=-C
\,\,\,\,\,,\,\,\,\,\,F=0\,,
\label{werr}
\end{equation}
where a prime means differentiation with respect to $\chi$.
The various $\psi$-kinetic terms in (\ref{skl}) arise from the demand that their total
covariant derivative cancels against the cosmological constant term $\Lambda g_{\mu\nu}$ and this
is enough in order to determine their form uniquely. The important point is that the covariant
cancelation of the $\chi$-kinetic terms occurs against zero, and therefore, a new arbitrary field $C$
arises. At the cosmological level that we elaborate it is $\chi=\chi(t)$, $C=C(t)$
and $C'=\dot{C}/\dot{\chi}$. Then, the consistency of equations (\ref{wkd}), (\ref{sdf})
will provide a differential equation of second order for $C(t)$. Indeed, the presence of the
$\chi$ terms in the field equations will add extra terms on the right hand sides of (\ref{wkd}),
(\ref{sdf}). Each such term contains products of $m$-th time derivatives of $C$ with $n$-th
time derivatives of $\chi$ ($m,n=0,1,2$), where let's denote such products as
$C^{(m)}\chi^{(n)}$ for convenience. The conservation equation (\ref{elk}) remains the same.
Following the same process which led to (\ref{lej}), instead of the constraint (\ref{constr}),
we will obtain a differential equation of second order for $C(t)$. This equation will still
contain $P+\frac{1}{3}\rho$ as one term, while all the others will be $C^{(m)}\chi^{(n)}$ terms.
If we choose the initial conditions $C(0)\simeq 0$, $\dot{C}(0)\simeq 0$ for the differential equation, then
in a short time interval around $t=0$, equation (\ref{constr}) will arise approximatively. At the
same time, in this interval the tensor $\theta_{\mu\nu}^{(G)}$ will be approximately
zero and equations (\ref{skl}) will arise. The result is that our inclusion of only the
$\Lambda$-kinetic terms, implying the constraint (\ref{constr}), simplifies the analysis
without mixing up with extra integration constants, and moreover is a consistent option at early times.

Let us finish with a few comments. First, a non-conservation equation of the form (\ref{elk})
implies an energy transfer between the energy density $\rho$ and the gravitational coupling $G$.
Consistently with the scaling (\ref{scaling}), it is $\frac{\dot{G}}{G}=-2\frac{\dot{k}}{k}=
-\frac{\dot{\Lambda}}{\Lambda}$,
and it will be verified from the following matter solutions that in most cases $k(t)$ decreases.
Thus, $\Lambda$ also decreases with time and there is an energy transfer from $\rho$ to $G$. In one
matter solution it is found that $k(t)$ increases and the opposite behaviour of a transfer from $G$ to
$\rho$ occurs. Moreover, the above equation can lead to entropy production and reheating,
as will be discussed in section \ref{inflation}.
Second, a constraint of the form (\ref{constr}) does not appear in other studies
of AS inspired cosmologies and it implies an extra negative pressure $\pi$ (whenever $\rho+3p>0$). Then,
from equation (\ref{elk}) it turns out that the decay rate of the energy density $\rho$ becomes smaller
compared to the free dilution. One way to interpret this negative pressure is due to particle
production and matter creation and another way is due to bulk viscosity (both mechanisms can be
present simultaneously, but this situation will not be considered here due to complexity).
Third, in the case of particle production, $\pi$ is solely given by equation (\ref{constr}) and
then, equations (\ref{wkd}), (\ref{elk}) provide the solution assuming the AS scaling (\ref{scaling})
as well as an energy-length scaling (this case will be examined in the subsection
\ref{particle production}).
The subcase $\pi=0$ has the same treatment and provides the cosmic string-like
equation of state $p=-\frac{1}{3}\rho$, which for a standard dilution of $\rho$ is also the equation
of state for the curvature term; this equation of state also arises out of dimensional arguments
assuming that the mass $M$ of a spherical region obeys in the early universe a Machian expression
where $G_{\!N}M$ is proportional to the radius $r$ \cite{modphys}.
In the case of bulk viscosity, $\pi$ is additionally given by another expression, and equations
(\ref{wkd}), (\ref{elk}), (\ref{constr}) provide the solution given the scaling properties
(\ref{scaling}), but with the difference that now the energy-length scale is determined by these
equations (this case will be examined in the subsection \ref{bulk viscosity}).

At the NGFP (\ref{scaling}), the conservation equation (\ref{elk}), together with the constraint
(\ref{constr}), is written as
\begin{equation}
\dot{\rho}=2\rho\Big(\frac{\dot{k}}{k}-H\Big)\,.
\label{kwh}
\end{equation}
Thus, depending on the sign of $\dot{k}$, the energy density $\rho$ can either decrease or increase.
Usually $k,\rho$ decrease with time, but
since in the present work it is proposed that the vacuum solutions probably describe the initial
stage of the universe, it can be allowed to have $k$ temporarily increasing in a subsequent stage
of matter solutions.

\subsection{Particle production}
\label{particle production}

As explained above, the model at hand possesses naturally a non-equilibrium pressure
$\pi$ given in terms of the energy density $\rho$ and pressure $p$ by the expression (\ref{constr})
\begin{equation}
\pi=-\Big(\frac{1}{3}\rho+p\Big)\,.
\label{wef}
\end{equation}
This pressure turns out to be negative (as long as $\rho+3p>0$).
Equation (\ref{wef}) is also written as
\begin{equation}
\pi=-\frac{\rho+p}{3H}\,\frac{\dot{N}}{N}=-\frac{\rho+p}{\textrm{n}}\,\frac{dN}{dV}\,,
\label{dnr}
\end{equation}
where the ratio of the change of the number $N$ of particles in the proper comoving volume
$V\propto a^{3}$ is
\begin{equation}
\frac{\dot{N}}{N}=\frac{\rho+3p}{\rho+p}H
\label{lwa}
\end{equation}
and $\textrm{n} =\frac{N}{V}$ is the particle number density. Due to the second equation in (\ref{dnr}),
the conservation equation (\ref{elk}) is written as
\begin{equation}
d(\rho V)+pdV-\frac{\rho+p}{\textrm{n}}dN+\frac{\rho V}{G} dG=0\,.
\label{egt}
\end{equation}
The third term in this equation expresses the presence of matter creation in the context of
open systems \cite{Prigogine:1989zz}, with the important difference that here the form of this
creation is predicted by the theory itself, as given by equation (\ref{lwa}).
Therefore, one way to interpret the supplementary pressure $\pi$ of equation (\ref{wef})
is that it corresponds to particle production. Equation (\ref{egt}) expresses the thermodynamical energy
conservation of an open system in the case of adiabatic transformation ($dQ=0$) and the ``heat''
exchanged by the system in our case is due not only to the change of the number of particles but
also to the change of the gravitational constant $G$. Equation (\ref{lwa}) is of a special form
among the various models in the literature parametrizing the particle change rate in the case of
isentropic particle production as $\frac{\dot{N}}{N}=3\beta H_{\ast}(\frac{H}{H_\ast})^{\alpha}$
\cite{freaza}, \cite{waga}, where $\alpha, \beta$ are
$\mathcal{O}(1)$ dimensionless constants and $H_{\ast}$ is a reference value,
e.g. the present Hubble rate.
So, for $p=w\rho$, in our case it is $\alpha=1$ and
$\beta=\frac{1+3w}{3(1+w)}$ (for example, for a reasonable equation of state in the early universe
that of relativistic matter $p=\frac{1}{3}\rho$ it is $\pi=-\frac{2}{3}\rho$).
Therefore, it is remarkable that the very same modified Einstein equations suggest a transfer of
energy from the gravitational field to matter through particle production.

Integration of the non-conservation equation (\ref{elk}), using (\ref{wef}), gives
\begin{equation}
\rho=\frac{\rho_{o}}{G a^{2}}\,,
\label{edv1}
\end{equation}
where $\rho_{o}>0$ is an integration constant (note that no particular $w$ has been chosen).
Finally, the Friedmann equation (\ref{wkd}) is written as
\begin{equation}
H^{2}+\frac{\kappa}{a^{2}}=\frac{\bar{\Lambda}}{3}e^{\psi}-H\dot{\psi}-
\frac{1}{4}\dot{\psi}^{2}+\frac{8\pi}{3}\frac{\rho_{o}}{a^{2}}\,.
\label{sdr}
\end{equation}
Note that $G$ has disappeared in (\ref{sdr}) and no particular form for $G(k)$ has been assumed.
However, according to AS, the forms $G=\frac{g_{\ast}}{\xi^{2}}t^{2}$ for the scaling
$k=\frac{\xi}{t}$, or $G=\frac{g_{\ast}}{\xi^{2}}H^{-2}$ for $k=\xi H$ are needed for the
determination of $\rho$ in (\ref{edv1}). Setting
\begin{equation}
\mu=\kappa-\frac{8\pi\rho_{o}}{3}\,,
\label{miu}
\end{equation}
equation (\ref{sdr}) is written as
\begin{equation}
H^{2}+\frac{\mu}{a^{2}}=\frac{\bar{\Lambda}}{3}e^{\psi}-H\dot{\psi}-\frac{1}{4}\dot{\psi}^{2}\,.
\label{qwr}
\end{equation}
If $\mu=0$, which means $\kappa=1$ and $\rho_{o}=\frac{3}{8\pi}$, equation (\ref{qwr}) is
identical with the vacuum equation (\ref{eqvac1}) with $\kappa_{v}=0$ (we denote by
$\kappa_{v}$ the curvature index of the vacuum case); thus the solutions in this
case coincide with the vacuum solutions of the previous section with $\kappa_{v}=0$.
Therefore, in this case there are strong power law inflationary solutions close to the initial
singularity.
\newline
If $\mu\neq 0$, equation (\ref{qwr}) takes the form
\begin{equation}
\frac{1}{a^{2}}\Big(\frac{da}{dt'}\Big)^{2}+\frac{\text{sgn}(\mu)}{a^{2}}=\frac{\bar{\Lambda}'}{3}
e^{\psi}-\frac{1}{a}\,\frac{da}{dt'}\,\frac{d\psi}{dt'}-\frac{1}{4}\Big(\frac{d\psi}{dt'}\Big)^{2}\,,
\label{egu}
\end{equation}
where $t'=\sqrt{|\mu|}\,t$, $\bar{\Lambda}'=\bar{\Lambda}/|\mu|$ and $\text{sgn}(\mu)$ denotes the
sign of $\mu$. Equation (\ref{egu}) coincides with the vacuum equation (\ref{eqvac1}) given that
$\text{sgn}(\mu)=\kappa_{v}\neq 0$ and $t'$ is replaced by $t$. So, for $\kappa=1$, $\rho_{o}<
\frac{3}{8\pi}$, the solutions coincide with the vacuum solutions with $\kappa_{v}=1$, just
rescaling time. Therefore, in this case there are accelerating (inflationary) solutions which either
avoid the big band singularity, or possess a big bang, or during a collapsing phase avoid the big
crunch towards expansion. Note from equation (\ref{edv1}) that when $a\neq 0$ and
$k_{\text{max}}<\infty$,
as happens with the non-singular vacuum solutions found previously, the energy density $\rho$
remains finite. For $\kappa=1$, $\rho_{o}>\frac{3}{8\pi}$ or for $\kappa\leq 0$ with
any $\rho_{o}$, the solutions coincide with the vacuum solutions with $\kappa_{v}=-1$,
just rescaling time. In this case a singular accelerating cosmology can occur. The
property of a decreasing $k(t)$ shown for the vacuum expanding solutions is also transferred
to the associated matter solutions discussed here. As for the energy density $\rho$, it decreases
with time due to (\ref{kwh}).

To summarize with the most interesting matter solutions with particle production, they refer
to the positively curved case and have power law inflation or are non-singular and accelerating.

\subsection{Bulk viscosity}
\label{bulk viscosity}

Since detailed physics in the proximity of the NGFP is still unknown, it is worth exploring the
possibility that the negative non-equilibrium pressure $\pi$ of equation (\ref{constr}) is due
to non-zero bulk viscosity through dissipative processes.
In this case the bulk viscous pressure $\pi$ has the form
\begin{equation}
\pi=-\zeta u^{\mu}{_{;\mu}}=-3\zeta H\,,
\label{viscosity}
\end{equation}
where $\zeta$ is the bulk viscosity coefficient, which will be assumed here to be constant.
Bulk pressures could be the consequence of the process where different matter components cool
with the expansion of the universe with different rates and the system moves away from
equilibrium. The expression (\ref{viscosity}) arises in some limit in the context of the
second-order theory of non-equilibrium thermodynamics \cite{Israel:1979wp}.
If we assume $\rho+3p>0$, for an expanding phase of the universe the expression (\ref{viscosity})
is consistent with the constraint (\ref{constr}) given that $\zeta>0$. This means that there are no
contracting parts in a solution at all, and all solutions are expanding. A reasonable equation
of state in the early universe is that of relativistic matter $p=\frac{1}{3}\rho$. However,
this restriction is not essential for the following analysis, so we assume a general barotropic
fluid with $p=w\rho$ and $1+3w>0$. Then, from equations (\ref{constr}), (\ref{viscosity}) it arises
a direct connection between the energy density and the Hubble parameter
\begin{equation}
\rho=\frac{9}{1\!+\!3w}\zeta H\,.
\label{wex}
\end{equation}
Integration of the non-conservation equation (\ref{elk}) gives
\begin{equation}
\rho=\frac{\rho_{o}}{G a^{2}}\,,
\label{edv}
\end{equation}
where $\rho_{o}>0$ is an integration constant.
The combination of (\ref{wex}), (\ref{edv}) gives
\begin{equation}
H=\frac{(1\!+\!3w)\rho_{o}}{9\zeta}\frac{1}{Ga^{2}}\,.
\label{sfr}
\end{equation}
Close to the NGFP defined by (\ref{scaling}) it arises from (\ref{sfr}) that
\begin{equation}
e^{\psi}=\nu H a^{2}\,\,\,\,\,,\,\,\,\,\,
\nu=\frac{9\zeta g_{\ast}\lambda_{\ast}}{(1\!+\!3w)\rho_{o}\bar{\Lambda}}>0\,.
\label{drh}
\end{equation}

Finally, the Friedmann equation (\ref{wkd}) gives due to (\ref{edv})
\begin{equation}
H^{2}+\frac{\kappa}{a^{2}}=\frac{\bar{\Lambda}}{3}e^{\psi}-H\dot{\psi}-
\frac{1}{4}\dot{\psi}^{2}+\frac{8\pi}{3}\frac{\rho_{o}}{a^{2}}\,.
\label{sdl}
\end{equation}
Plugging (\ref{drh}) into (\ref{sdl}) and converting the time derivatives to $a$-derivatives
we get the equation
\begin{equation}
a^{2}\Big(\frac{dH}{da}\Big)^{2}+8aH\frac{dH}{da}+16H^{2}-\frac{4\nu\bar{\Lambda}}{3}
a^{2}H+4\Big(\kappa\!-\!\frac{8\pi\rho_{o}}{3}\Big)\frac{1}{a^{2}}=0\,.
\label{lwf}
\end{equation}
Setting
\begin{equation}
z=a^{4}H>0\,\,\,\,\,\,\,,\,\,\,\,\,\,\,x=a^{3}\,,
\label{krg}
\end{equation}
equation (\ref{lwf}) gets the form
\begin{equation}
\frac{dz}{dx}=\pm\sqrt{\frac{4\nu\bar{\Lambda}}{27}z\!+\!\frac{4}{9}
\Big(\frac{8\pi\rho_{o}}{3}\!-\!\kappa\Big)}\,,
\label{skr}
\end{equation}
where the square root has to be positive. Integration of (\ref{skr}) gives
\begin{equation}
H=\beta a^{2}\pm\frac{c}{a}+\frac{\gamma}{a^{4}}\,,
\label{teo}
\end{equation}
where $c$ is integration constant and
\begin{equation}
\beta=\frac{\nu\bar{\Lambda}}{27}>0\,\,\,\,\,\,\,,\,\,\,\,\,\,\,
\gamma=\frac{27}{4\nu\bar{\Lambda}}\Big[c^{2}-\frac{4}{9}\Big(\frac{8\pi\rho_{0}}{3}\!-\!\kappa\Big)
\Big]\,,
\label{eko}
\end{equation}
under the constraints $c\pm 2\beta a^{3}>0$, $\beta a^{6}\pm ca^{3}+\gamma>0$. For
$y=c\pm 2\beta a^{3}$, the first of these constraints become $y>0$ and the second
$y^{2}>c^{2}-4\beta\gamma=\frac{4}{9}(\frac{8\pi\rho_{o}}{3}-\kappa)$.
\begin{itemize}
\item{
If $c^{2}-4\beta\gamma<0
\Leftrightarrow \kappa=1$, $\rho_{o}<\frac{3}{8\pi}$, the only constraint is
$c\pm 2\beta a^{3}>0$ and there are three cases: (i) for the upper
branch with $c<0$ it is $a>(\frac{|c|}{2\beta})^{1/3}$, (ii) for the upper branch with $c>0$ there
is no bound on $a$, and (iii) for the lower branch it is $c>0$ and $a<(\frac{c}{2\beta})^{1/3}$.}
\item{
If $c^{2}-4\beta\gamma>0\Leftrightarrow \kappa\leq 0$, or $\kappa=1$, $\rho_{o}>\frac{3}{8\pi}$,
the only constraint is $c\pm 2\beta a^{3}>\sqrt{c^{2}-4\beta\gamma}$.
For the upper branch there are two cases: (i) if $c<0$, or if $c>0$, $\gamma<0$ it is
$a>[\frac{1}{2\beta}(\sqrt{c^{2}-4\beta\gamma}-c)]^{1/3}$ and (ii) if $c>0$, $\gamma>0$ there is no
bound on $a$. For the lower branch it has to be $c>0$, $\gamma>0$ and
$a<[\frac{1}{2\beta}(c-\sqrt{c^{2}-4\beta\gamma})]^{1/3}$.}

\end{itemize}

From equations (\ref{wex}), (\ref{teo}) it is obvious that the energy density is finite for the
solutions which avoid the zero scale factor, so the universe avoids the infinite density singularity.
Moreover, from equation (\ref{teo}) we can calculate the Ricci scalar, which takes the form
\begin{equation}
\frac{R}{6}=4\beta^{2}a^{4}\pm 5\beta c a+\frac{2\beta\gamma\!+\!c^{2}\!+\!\kappa}{a^{2}}
\mp\frac{\gamma c}{a^{5}}-\frac{2\gamma^{2}}{a^{8}}\,.
\label{jef}
\end{equation}
Therefore, the solutions which avoid a infinite density singularity avoid also a curvature
singularity. Concerning the acceleration it is found similarly
\begin{equation}
\frac{\ddot{a}}{3a^{2}}=\beta (\beta a^{3}\pm c)-\frac{\gamma}{a^{9}}(\gamma\pm ca^{3})\,.
\label{ole}
\end{equation}
For the case (i) above with $c^{2}-4\beta\gamma<0$ the universe at its minimum scale factor
starts decelerating and enters into acceleration. For the case (i) with $c^{2}-4\beta\gamma>0$
the universe at its minimum scale factor starts with zero acceleration, and immediately after, it
accelerates.

We summarize saying that there are branches of solutions for any
spatial topology which are expanding, non-singular and accelerating.

The dependence of the scale factor with time can be found integrating equation (\ref{teo})
\begin{eqnarray}
t-t_{0}&=&\int\frac{da}{\beta a^{3}\pm c+\gamma a^{-3}}
\label{rlh}\\
&=&\frac{1}{3\beta}\int\frac{(u+\sigma)^{\frac{1}{3}}}{u^{2}\!-\!\tau}du\,,
\label{huw}
\end{eqnarray}
where $t_{0}$ is integration constant and
\begin{equation}
u=a^{3}\pm\frac{c}{2\beta}\,\,\,\,\,\,,\,\,\,\,\,\,\sigma=\mp\frac{c}{2\beta}\,\,\,\,\,\,,
\,\,\,\,\,\,\tau=\frac{c^{2}\!-\!4\beta\gamma}{4\beta^{2}}\,.
\label{eth}
\end{equation}
In the case $c^{2}\!-\!4\beta\gamma>0$, this integral can be performed analytically in closed form.
First, it is
\begin{equation}
6\beta\sqrt{\tau}\,(t-t_{0})=
\theta^{\frac{1}{3}}\!\int\frac{v^{\frac{1}{3}}}{v\!-\!\epsilon}dv-
\tilde{\theta}^{\frac{1}{3}}\!\int\frac{\tilde{v}^{\frac{1}{3}}}
{\tilde{v}\!-\!\tilde{\epsilon}}d\tilde{v}\,,
\label{rlg}
\end{equation}
where $v=\theta^{-1}a^{3}$, $\tilde{v}=\tilde{\theta}^{-1}a^{3}$,
$\theta=|\sigma\!+\!\sqrt{\tau}|$, $\tilde{\theta}=|\sigma\!-\!\sqrt{\tau}|$, $\epsilon=\text{sgn}
(\sigma\!+\!\sqrt{\tau})$, $\tilde{\epsilon}=\text{sgn}(\sigma\!-\!\sqrt{\tau})$. We write
\begin{eqnarray}
2\epsilon\int\frac{v^{\frac{1}{3}}}{v\!-\!\epsilon}dv
&=&\int\frac{1\!+\!2\epsilon v^{\frac{1}{3}}}{v\!-\!\epsilon}
dv-\int\frac{1}{v\!-\!\epsilon}dv=
\int\frac{1}{v^{\frac{1}{3}}\!-\!\epsilon}dv
\!-\!\int\frac{v^{\frac{1}{3}}}{v^{\frac{2}{3}}\!+\!\epsilon
v^{\frac{1}{3}}+1}dv-\int\frac{1}{v\!-\!\epsilon}dv\\
&=&3\int\frac{q^{2}}{q\!-\!\epsilon}dq-3\int\frac{q^{3}}{q^{2}\!+\!\epsilon q\!+\!1}dq
-\ln|v-\epsilon|\,\,\,\,\,\,\,,\,\,\,\,\,\,\,q=v^{\frac{1}{3}}=\theta^{-\frac{1}{3}}a\\
&=&6\epsilon v^{\frac{1}{3}}+\ln\frac{|v^{\frac{1}{3}}\!-\!\epsilon|^{3}}{|v\!-\!\epsilon|}
-2\sqrt{3}\,\epsilon\arctan\frac{2v^{\frac{1}{3}}\!+\!\epsilon}{\sqrt{3}}\,.
\end{eqnarray}
Finally,
\begin{equation}
6\beta\sqrt{\tau}\,(t-t_{0})=\frac{1}{2}\ln\!
\left[\left(\frac{|\theta^{-\frac{1}{3}}a\!-\!\epsilon|^{3}}
{|\theta^{-1}a^{3}\!-\!\epsilon|}\right)^{\!\!\epsilon\theta^{1/3}}
\!\!\left(\frac{|\tilde{\theta}^{-1}a^{3}
\!-\!\tilde{\epsilon}|}{|\tilde{\theta}^{-\frac{1}{3}}a\!-\!\tilde{\epsilon}|^{3}}
\right)^{\!\!\tilde{\epsilon}
\tilde{\theta}^{1/3}}\right]-\sqrt{3}\Bigg(\!\theta^{\frac{1}{3}}\arctan
\frac{2\theta^{-\frac{1}{3}}a\!+\!\epsilon}{\sqrt{3}}-\tilde{\theta}^{\frac{1}{3}}\arctan
\frac{2\tilde{\theta}^{-\frac{1}{3}}a\!+\!\tilde{\epsilon}}{\sqrt{3}}\Bigg)
\,.
\label{rei}
\end{equation}

Note that in the present case of bulk viscosity, it was nowhere assumed some energy-length scaling.
Actually $k(t)$ is determined from equation (\ref{drh}) as follows
\begin{equation}
k=\sqrt{\frac{9\zeta g_{\ast}}{(1\!+\!3w)\rho_{o}}}\,\,a\,\sqrt{H}\,.
\label{kkk}
\end{equation}
For small scale factors with $\gamma>0$, due to (\ref{teo}), equation (\ref{kkk}) implies
$k\sim\frac{1}{a}$, so $k$ scales inversely proportional to the proper distance at fixed time.
It is worth emphasizing that here the physics of the fluid is this that determines the function
$k(t)$. This property is reasonable since different matter content should necessarily result to
different scaling laws due to concrete physical reasons. Indeed, in reality the details of the
``thermodynamic'' (or the essential relevant parameters in case of non-equilibrium evolution)
properties of the statistical ensemble of quantum particles should determine how ``strong'' the
relation is between the measure of mean energy $k$ and the geometrical cosmological measure of the
``distance''.
For the solution with $c^{2}-4\beta\gamma<0$ which possesses a minimum scale factor, it can be shown
that the scale $k(t)$ decreases in a region near the minimum (and also $\rho$ decreases).
However, for larger values of $a$ the function $k(t)$ increases.
For the non-singular solution with $c^{2}-4\beta\gamma>0$ the
function $k(t)$ is found to increase near the minimum scale factor (and also $\rho$ increases).
Both cases with increased $k$ can be interpreted as intermediate stages in the cosmic evolution.

\section{Inflation, Reheating and Entropy generation}
\label{inflation}

Here, a short discussion about the inflationary period and the possible subsequent reheating
and entropy production will be given.
Contrary to the aim of several other works concerning global cosmological solutions in the AS program,
the focus here is the cosmic period near the NGFP regime. This high energy regime is of particular
importance for the possibility of an inflationary period. Fortunately, this is also a regime where
the behaviour of $\Lambda$ and $G$ is known much better and there are geometry independent methods
handling the running of these couplings. The solutions found in the previous sections possess
accelerating phases either in the vacuum or the matter sector.
The cosmic scenario we are going to analyze is based on the
assumption that the universe first starts in a pure vacuum (perhaps creation of the universe from a
vacuum fluctuation), where the relevant equations of motion contain only a vacuum contribution.
Subsequently, it enters a period where matter starts to become more important (still inside the
NGFP regime) that ends when $k\thickapprox m_{pl}$. At this energy scale there is a transition towards
a third stage of
conventional FRW universe with negligible $\Lambda$ and constant $G$. The derived solutions in the
preceding sections are able to model both the first two stage cosmic evolution. The second stage can
be modeled either by solutions that suggest, as it will be seen, particle production with entropy generation
or by solutions with bulk viscosity associated with entropy production and possible reheating.
Both these matter solutions are described by equations which are consistent with the
vacuum case equations.

\subsection{First stage: Inflationary cosmogenesis}
\label{first stage}

An acceptable approximation \cite{2007JCAP...08..024B} to describe the RG improved UV early
cosmological history is to work separately at the different three stages using the developed solutions.
It remains to the details of a full RG running to prove that the derived classical
cosmological solutions (used to describe the first two stages of cosmic evolution), inspired by the
energy scaling of the couplings $\Lambda$ and $G$, are indeed fair approximations of the quantum
average spacetime that describes the early universe.

Some long standing qualitative arguments speculate that due to Heisenberg's uncertainty principle,
universe was created from ``empty'' spacetime. A small true vacuum bubble/void of expanding vacuum
space, can be created probabilistically by quantum fluctuations of a metastable false vacuum through
a first/second order phase transition. If this initial bubble/void cannot expand rapidly, it will
disappear soon. In case this initial baby universe expands rapidly to a large enough size, the
universe can then be created irreversibly.
This baby universe created probabilistically by quantum vacuum
fluctuations starts with a finite volume. Thus, it is expected that the energy scale $k$ may not
initiate from infinity and the corresponding $\Lambda(k)$ is finite.

In more detail we consider a time interval $t_0<t< t_{1}$, where $t_0$ is the initial time
of quantum birth and $t_{1}$ is the transition time to the second stage of matter appearance.
From the derived vacuum solutions, we will pick the simple power-law spatially flat solution
(\ref{hgr}) to model this era,
\begin{equation}
a(t)=a_0\Big( \frac{t}{t_0} \Big)^{1+\omega}\,\,\,\,\,\,,\,\,\,\,\,\,
\text{for} \,\,\,\, t\geq t_0\,,
\label{first}
\end{equation}
where $a_{0}$ is the initial scale  factor at $t_{0}$. The larger the value of $\omega$,
the stronger the inflation is. For a structure of comoving length $\Delta x$, the corresponding
physical (proper) length at any $t$ is $L(t)=a(t)\Delta x$. Due to equation (\ref{first}) it is
\begin{equation}
\label{infl}
L(t)= \Big(\frac{t}{\,t_{1} }\Big)^{1+\omega}L(t_{1})\,.
\end{equation}
Now, the Hubble radius $\ell_{H}(t)\equiv\frac{1}{H(t)}$ is given by
\begin{equation}
\label{lhub}
\ell_{H}(t)=\frac{t}{1+\omega}\,.
\end{equation}
In order to study when $L(t)$ crosses the Hubble radius $\ell_{H}(t)$ we evaluate their ratio
\begin{equation}
\label{crossing}
\frac{L(t)}{\ell_{H}(t)}=\Big(\frac{t}{t_{1}}\Big)^{\omega} \,\frac{L(t_{1})}{\ell_{H}(t_{1})}\,.
\end{equation}
It is obvious that the proper length of a part of the universe increases
fast enough to cross the Hubble radius.
The desired $60$ e-folds can be easily achieved for moderate values of $\omega$. Indeed,
let us assume for simplicity that all the required $60$ e-foldings are achieved during the first
cosmological stage, although it is possible to have a second inflationary period with different
characteristics during the second stage. Then, at $t=t_{1}$ we need $L(t_{1})$ to be
$e^{60}$ times the Hubble radius $\ell_{H}(t_{1})$ and we get
\begin{equation}
\label{tr1}
\frac{L(t)}{\ell_{H}(t)} = e^{60}\Big ( \frac{t}{t_{1}} \Big )^{\omega}\,.
\end{equation}
The time when $L$ crosses the Hubble radius happens for $t=t_{cr}$ with
$L(t_{cr})=\ell_{H}(t_{cr})$, and equation (\ref{tr1}) becomes
\begin{equation}
\label{tcross}
t_{cr}=t_{1}\,e^{-\frac{60}{\omega}}\,.
\end{equation}
It is obvious from the above equation that for moderate values of $\omega$, the time $t_{cr}$
can be much shorter that the transition time $t_{1}$.

\subsection{Second stage: Heat transfer and Entropy production}
\label{second stage}

Subsequently, a second cosmic period holds for $t_{1}<t< t_{2}$, where $t_{2}$ is the transition
to FRW universe. Here, apart from the vacuum contribution there is also matter. The study of the
matter solutions derived previously reveals the existence of either deceleration or inflationary
eras. Since now matter is present, it is essential to analyze the thermodynamics of the universe.

\textit{Entropy production through particle production.}
In the case of particle production,
thermodynamics of open systems, as applied to cosmology, takes into account both matter and entropy
creation on a macroscopical level. This consideration generalizes the standard thermodynamics
in cosmology, since beyond $\rho$ and $p$, the particle density $\textrm{n}$ also enters naturally.
If $U=\rho V$ is the internal energy in a proper comoving volume $V$ with corresponding entropy $S$
and temperature $T$, the entropy change $dS$ is given by
\begin{equation}
TdS=d(\rho V)+pdV-\mu dN=\frac{\rho+p}{\textrm{n}}dN-\mu dN-\frac{\rho V}{G} dG
=T\frac{s}{\textrm{n}}dN-\frac{\rho V}{G}dG\,.
\label{ker}
\end{equation}
The second equation arises do to (\ref{egt}) and
the third equation arises due to that the chemical potential $\mu$ is given by the Euler's
equation $\mu \textrm{n}=\rho+p-Ts$, where $s=\frac{S}{V}$ is the entropy per unit volume. As long as
the right hand side of equation (\ref{ker}) is positive, the second law of thermodynamics is satisfied,
$dS>0$. Using (\ref{lwa}), equation (\ref{ker}) reduces to a differential equation for the entropy $S$
\begin{equation}
\dot{S}=\frac{\rho+3p}{\rho+p}HS-\frac{\rho V}{T}\,\frac{\dot{G}}{G}\,.
\label{erg}
\end{equation}
It is not an easy issue \cite{2007JCAP...08..024B} to succeed at the same time entropy and
particle production and in the present work we have managed this.
Assuming a radiation equation of state, $w=\frac{1}{3}$, the Boltzmann law $\rho=\sigma_{B}T^{4}$
holds, and equation (\ref{erg}), due to (\ref{edv1}), takes the form
\begin{equation}
\frac{dS}{d\alpha}=\frac{3}{2}S+\frac{4}{3}v\sigma_{B}^{\frac{1}{4}}\rho_{o}^{\frac{3}{4}}
e^{\frac{3\alpha}{2}}\frac{dG^{-\frac{3}{4}}}{d\alpha}\,,
\label{awt}
\end{equation}
where $\alpha=\ln{a}$ and $V=va^{3}$ with $v$ being the comoving volume.
Integration of (\ref{awt}) gives the solution
\begin{equation}
S=va^{\frac{3}{2}}\Big(c+\frac{4\sigma_{B}^{\frac{1}{4}}\rho_{o}^{\frac{3}{4}}}
{3G^{\frac{3}{4}}}\Big)\,,
\label{wrf}
\end{equation}
where $c$ is integration constant. Using the NGFP scaling (\ref{scaling}) of $G$, we find
\begin{equation}
S=va^{\frac{3}{2}}\big(c+\nu k^{\frac{3}{2}}\big)\,,
\label{jee}
\end{equation}
where $\nu=\frac{4\sigma_{B}^{1/4}\rho_{o}^{3/4}}{3g_{\ast}^{3/4}}$.
From (\ref{jee}) it arises
\begin{equation}
\dot{S}=\frac{3}{2}va^{\frac{3}{2}}\Big[cH+\nu k^{\frac{3}{2}}\Big(H+\frac{\dot{k}}{k}\Big)\Big]\,.
\label{ewe}
\end{equation}
For the case of particle production we adopted two energy-length scalings.
For the first one, $k=\frac{\xi}{t}$, it is
\begin{equation}
\dot{S}=\frac{3}{2}va^{\frac{3}{2}}\Big[cH+\nu k^{\frac{3}{2}}\Big(H-\frac{1}{t}\Big)\Big]\,.
\label{won}
\end{equation}
It can be easily seen that all the corresponding interesting matter solutions with particle production
for any spatial topology $\kappa$ have $H-\frac{1}{t}>0$ in the expanding phase (this property
applies also for the non-singular solution found). So, for $c\geq 0$ there is a natural entropy
production. For the second scaling, $k=\xi H$, it is
\begin{equation}
\dot{S}=\frac{3}{2}va^{\frac{3}{2}}\Big(cH+\nu k^{\frac{3}{2}}\frac{\ddot{a}}{aH}\Big)\,.
\label{rlq}
\end{equation}
Therefore, if $c\geq 0$, whenever there is acceleration, at the same time there is an entropy
production. Now, the accelerating solutions found previously, with either $\kappa=0$ or
$\kappa=1$, share this property (the non-singular solution is included).

\textit{Entropy production through bulk viscosity.}
In the case of bulk viscosity, one can
use the standard thermodynamic relation of closed systems $dU+pdV=TdS$, from where
it arises immediately
\begin{equation}
\frac{T}{V}\dot{S}=\dot{\rho}+3H(\rho+p)\,.
\label{wro}
\end{equation}
Making use of the conservation equation (\ref{elk}) it turns out
\begin{equation}
\frac{T}{V}\dot{S}=-\rho\frac{\dot{G}}{G}-3H\pi\,.
\label{ory}
\end{equation}
In conventional FRW cosmology where the right hand side of equation (\ref{ory}) is zero, it can be
concluded that the entropy of a comoving volume remains the same as the universe expands, $\dot{S}=0$.
In our model, in the NGFP regime it is $\frac{\dot{G}}{G}=-2\frac{\dot{k}}{k}$, and since
usually $k$ drops with the expansion, in order to have increasing entropy we need
$\pi<0$ which is offered by the mechanism of bulk viscosity.
Moreover, in the study of bulk viscosity, we found previously one case where $k(t)$ increase, and thus,
$S(t)$ also increases.
\newline
Equation (\ref{ory}), making use of (\ref{constr}) with $w=\frac{1}{3}$, takes the form
\begin{equation}
\frac{T}{V}\dot{S}=\rho\Big(2H-\frac{\dot{G}}{G}\Big)=2\rho\Big(H+\frac{\dot{k}}{k}\Big)\,.
\label{iet}
\end{equation}
Finally, equation (\ref{iet}), using (\ref{kkk}), gives
\begin{equation}
\frac{T}{V}\dot{S}=\rho\Big(\frac{\dot{H}}{H}+4H\Big)=\frac{\rho}{H}\Big(
\frac{\ddot{a}}{a}+3H^{2}\Big)\,.
\label{rlv}
\end{equation}
It is possible to prove that all the upper branch solutions (\ref{teo}) are associated with entropy
increase (the non-singular solution is included).

\textit{Reheating.} Now, we are going to discuss the evolution of the temperature.
If we consider a radiation equation of state, the Boltzmann law of radiation is
$\rho=\sigma_{B} T^{4}$, and it follows from (\ref{kwh}) that
\begin{equation}
\dot{T}=\frac{T}{2}\Big(\frac{\dot{k}}{k}-H\Big)\,.
\label{kcj}
\end{equation}
Therefore, in order to have reheating, $\dot{T}>0$, the time derivative $\dot{k}$ has to be
sufficiently positive. For the particle production this is not true, since we have found that
$k(t)$ is permanently decreasing; in this case reheating could be realized by other
means, perhaps with the occurrence of possible phase transitions.
For the bulk viscosity, equation (\ref{kcj}) takes the form
\begin{equation}
\dot{T}=\frac{T}{4H}\dot{H}\,.
\label{nwj}
\end{equation}
Thus, in order to have a temperature raise, a superacceleration, $\dot{H}>0$, should occur.
For the non-singular solutions it can be shown that,
depending on the parameters, the universe can have $\dot{T}>0$ already from its minimum scale
factor onwards, or the temperature raise can appear later.


In summary, the first period cosmic evolution can be described by vacuum AS modified equations with
$\Lambda$ present. This period of cosmic genesis is associated with strong inflation. At the second
period the universe evolution is described by modified equations which include matter and is able to
solve the cosmological entropy problem. In the case of bulk viscosity there is also heat transfer
from vacuum to the matter sector.
Finally, the third stage of cosmic evolution, which is not described by the set of modified
Einstein equations presented here, happens when $k$ departs from $m_{pl}$. At this point another
semiclassical description of the spacetime applies where both $G$, $\Lambda$ are almost constant.
The framework of the AS program ensures that for lower energies conventional FRW
universe is recovered with negligible $\Lambda$.

\section{Discussion and Conclusions}
\label{conclusions}

General branches of new cosmological solutions have been obtained in the context of
Asymptotic Safe gravity (Quantum Einstein Gravity) at high energies close to the NGFP.
The derived solutions arise from a new consistent system of modified Einstein equations.
The framework handles two different cosmic periods. First a possible quantum birth from
a vacuum state and second the addition of a matter component at high energies.
However, the presented framework has to be modified to treat physics away from the transplanckian regime.

In the first cosmic era the source is an energy-dependent cosmological constant that scales at high
energies as the AS scenario suggests near the NGFP, i.e.
$\Lambda(k)\propto k^2$. This scaling is also the unique one which is consistent with dimensional
analysis without the introduction of a new energy scale. The cosmological constant becomes
time-dependent under the assumption of an energy-length scaling. The modified Einstein equations are
uniquely defined and arise by adding appropriate covariant kinetic terms of $\Lambda$ in
order to ensure the satisfaction of the Bianchi identities. The importance of
the presented vacuum solutions, consistent with a quantum birth, lies on the fact that they provide
inflationary expansion and at the same time completely remove the initial singularity in all
scale factor, energy density and curvature invariants. Exit from inflation is a natural output
of AS scenario and occurs when the energy scale becomes lower, and then, $\Lambda(k)$ becomes
insignificant and standard decelerating cosmology arises.

In the second cosmic era the inclusion of matter close to the NGFP was possible to be modeled
generalizing the vacuum equations of motion. An energy exchange arises between the matter and the
varying gravitational constant $G(k)\propto k^{-2}$. A negative non-equilibrium pressure beyond
the thermodynamic one is also an outcome and can be attributed to either a particle production or
to a mechanism of bulk viscosity. In both cases, there are general solutions which are
inflationary (with different characteristics than those arising during the first period)
and non-singular, and such behaviours can be found for any spatial topology. The barotropic
equation of state is not particularly significant.
In the case of bulk viscosity the relation between the energy scale and the time is implied
by the theory itself.
The most interesting feature of the matter solutions is that they suggest either particle production with
entropy generation or bulk viscosity with entropy production and reheating.

Since the presented solutions cover two consecutive cosmic eras, both close to the NGFP, various
phenomenological investigations worth to be investigated. It would certainly be interesting also
to study with the help of RG flow equations the transition between the two eras. Extending/generalizing
appropriately the present framework of modified Einstein equations and the associated energy conservation,
it may also be possible to describe the subplanckian cosmic evolution with emerging high energy corrections
to the conventional expansion rate that could explain baryogenesis \cite{Aliferis:2014ofa}, or dark energy
\cite{Kofinas:2011pq}.

Let us close with a few general comments. First, note that physical predictions, e.g. on the early
universe, should actually depend on universal quantities, like the critical exponent at a fixed point,
but not the fixed-point values themselves. This means that all the presented cosmological
solutions (vacuum and matter) are phenomenological in the same sense as the standard model
of Weinberg-Salam is. Although the formalism and all the derived solutions are general and
do not depend on specific values of the parameters $g_{\ast}$ and $\lambda_{\ast}$,
at the end, the analysis of the results depends on these parameters through inequalities
in the parameter space and not through fine-tuning. For example, the non-singular or the
inflationary behavior are not properties that arise from specific values of the parameters.
Only when the asymptotic safety program will be able to provide
the final Lagrangian and critical exponents \cite{Braun:2010tt}, precision conclusions
about the characteristics of the inflation will be possible. Note, however, that the final
picture regarding the understanding of the gravitational field would be much more conceptually
different than the case of the rest fermion/bosonic content \cite{Braun:2010tt}. The reason
is that we have to answer how measurements are performed, something that results on radical
new physical requirements. Thus, the presented solutions can be regarded as useful
phenomenological models of metrics that would probably describe a quantum gravity
inspired prototype model of the average spacetime near the NGFP, or possibly describe some
state spacetimes of the quantum ensemble. Nevertheless, it is natural and not problematic
(in the context of AS) to expect that near the NGFP there is a quantum superposition of
$\emph{non-singular}$ spacetimes.

\begin{acknowledgments}
V. Zarikas acknowledges the hospitality of Nazarbayev University.
\end{acknowledgments}


\begin{thebibliography}{99}

\bibitem{perimeter}
http://www.perimeterinstitute.ca/video-library/collection/renormalization-group-approaches-quantum-gravity.


\bibitem{reviews}
M.~Niedermaier and M.~Reuter,
 Living Rev.\ Rel.\  {\bf 9}, 5 (2006);
 M.~Reuter and F.~Saueressig,
 [arXiv:0708.1317 [hep-th]];
 R.~Percacci,
 In {\textit{Oriti, D. (ed.): Approaches to quantum gravity}} 111-128,
 [arXiv:0709.3851 [hep-th]];
 O.~Lauscher and M.~Reuter,
 In {\textit{Fauser, B. (ed.) et al.: Quantum gravity}} 293-313, [hep-th/0511260];
 M.~Reuter and F.~Saueressig,
 New J.\ Phys.\  {\bf 14}, 055022 (2012),
 [arXiv:1202.2274 [hep-th]];
 B.~Koch and F.~Saueressig,
 Int.\ J.\ Mod.\ Phys.\ A {\bf 29}, no. 8, 1430011 (2014),
 [arXiv:1401.4452 [hep-th]];
 A.~Bonanno,
 PoS CLAQG {\bf 08}, 008 (2011), [arXiv:0911.2727 [hep-th]];
 M.~Niedermaier,
 Class.\ Quant.\ Grav.\  {\bf 24}, R171 (2007), [gr-qc/0610018].


\bibitem{wein}
S.~Weinberg,
 In {\textit{Boston 1996, Conceptual foundations of quantum field theory}} 241-251, [hep-th/9702027];
 S.~Weinberg,
 PoS CD {\bf 09}, 001 (2009), [arXiv:0908.1964 [hep-th]].



\bibitem{mr}
M.~Reuter,
 Phys.\ Rev.\ D {\bf 57}, 971 (1998), [hep-th/9605030].



\bibitem{Wetterich:1992yh}
C.~Wetterich,
 Phys.\ Lett.\ B {\bf 301}, 90 (1993);
 T.~R.~Morris,
 Int.\ J.\ Mod.\ Phys.\ A {\bf 9}, 2411 (1994), [hep-ph/9308265];
 D.~Dou and R.~Percacci,
 Class.\ Quant.\ Grav.\  {\bf 15}, 3449 (1998), [hep-th/9707239];
 M.~Reuter and F.~Saueressig,
 Phys.\ Rev.\ D {\bf 65}, 065016 (2002), [hep-th/0110054];
 O.~Lauscher and M.~Reuter,
 Phys.\ Rev.\ D {\bf 66}, 025026 (2002), [hep-th/0205062];
 W.~Souma,
 Prog.\ Theor.\ Phys.\  {\bf 102}, 181 (1999), [hep-th/9907027];
 A.~Bonanno and M.~Reuter,
 JHEP {\bf 0502}, 035 (2005), [hep-th/0410191];
 R.~Percacci and D.~Perini,
 Phys.\ Rev.\ D {\bf 68}, 044018 (2003), [hep-th/0304222].



\bibitem{Litim}
 D.~F.~Litim,
 Phys.\ Rev.\ Lett.\  {\bf 92}, 201301 (2004), [hep-th/0312114];
 P.~Fischer and D.~F.~Litim,
 Phys.\ Lett.\ B {\bf 638}, 497 (2006), [hep-th/0602203];
 J.~E.~Daum and M.~Reuter,
 Adv.\ Sci.\ Lett.\  {\bf 2} (2009) 255, [arXiv:0806.3907 [hep-th]];
 D.~Benedetti, K.~Groh, P.~F.~Machado and F.~Saueressig,
 JHEP {\bf 1106}, 079 (2011), [arXiv:1012.3081 [hep-th]].



\bibitem{rev2}
M.~Reuter and H.~Weyer,
 Gen.\ Rel.\ Grav.\  {\bf 41}, 983 (2009), [arXiv:0903.2971 [hep-th]];
 M.~Reuter and H.~Weyer,
Phys.\ Rev.\ D {\bf 79}, 105005 (2009), [arXiv:0801.3287 [hep-th]];
 P.~F.~Machado and R.~Percacci,
 Phys.\ Rev.\ D {\bf 80}, 024020 (2009), [arXiv:0904.2510 [hep-th]];
 E.~Manrique and M.~Reuter,
 PoS CLAQG {\bf 08}, 001 (2011), [arXiv:0905.4220 [hep-th]].


\bibitem{2006PhRvL..97v1301C}
 A.~Codello and R.~Percacci,
 Phys.\ Rev.\ Lett.\  {\bf 97}, 221301 (2006), [hep-th/0607128].


\bibitem{bms1}
A.~Bonanno,
 Phys.\ Rev.\ D {\bf 85}, 081503 (2012), [arXiv:1203.1962 [hep-th]].


\bibitem{bms2}
D.~Benedetti, P.~F.~Machado and F.~Saueressig,
 AIP Conf.\ Proc.\  {\bf 1196}, 44 (2009), [arXiv:0909.3265 [hep-th]].



\bibitem{Niedermaier:2009zz}
M.~Niedermaier,
 Nucl.\ Phys.\ B {\bf 833}, 226 (2010).


\bibitem{cpr1}
A.~Codello, R.~Percacci and C.~Rahmede,
 Int.\ J.\ Mod.\ Phys.\ A {\bf 23}, 143 (2008), [arXiv:0705.1769 [hep-th]];
A.~Codello, R.~Percacci and C.~Rahmede,
 Annals Phys.\  {\bf 324} (2009) 414, [arXiv:0805.2909 [hep-th]].


\bibitem{ms1}
P.~F.~Machado and F.~Saueressig,
 Phys.\ Rev.\ D {\bf 77}, 124045 (2008), [arXiv:0712.0445 [hep-th]].


\bibitem{Bonanno:1998ye}
A.~Bonanno and M.~Reuter,
 Phys.\ Rev.\ D {\bf 60}, 084011 (1999), [gr-qc/9811026];
A.~Bonanno and M.~Reuter,
 Phys.\ Rev.\ D {\bf 62}, 043008 (2000), [hep-th/0002196].


\bibitem{Bonanno:2006eu}
A.~Bonanno and M.~Reuter,
 Phys.\ Rev.\ D {\bf 73} (2006) 083005, [hep-th/0602159].



\bibitem{Reuter:2006rg}
M.~Reuter and E.~Tuiran, The Eleventh Marcel Grossmann Meeting, September 2008, 2608-2610,
 [hep-th/0612037];
M.~Reuter and E.~Tuiran,
 Phys.\ Rev.\ D {\bf 83} (2011) 044041, [hep-th/1009.3528].



\bibitem{Falls:2012nd}
K.~Falls and D.~F.~Litim,
 Phys.\ Rev.\ D {\bf 89} (2014) 084002, [gr-qc/1212.1821].


\bibitem{Cai:2010zh}
Y.-F.~Cai and D.~A.~Easson,
 JCAP {\bf 1009} (2010) 002, [hep-th/1007.1317].



\bibitem{Becker:2012js}
D.~Becker and M.~Reuter,
 JHEP {\bf 1207} (2012) 172, [hep-th/1205.3583];
D.~Becker and M.~Reuter, 13th Marcel Grossmann Meeting on Recent Developments in Theoretical and Experimental General Relativity, Astrophysics, and Relativistic Field Theories (MG13), p.2230-2232 Proceedings, [hep-th/1212.4274].


\bibitem{Ward:2006vw}
B.~F.~L.~Ward,
 Acta Phys.\ Polon.\ B {\bf 37} (2006) 1967, [hep-ph/0605054].



\bibitem{Basu:2010nf}
S.~Basu and D.~Mattingly, Phys. Rev. D 82 (2010) 124017, [hep-th/1006.0718].

\bibitem{Casadio:2010fw}
R.~Casadio, S.~D.~H.~Hsu and B.~Mirza, Phys. Lett. B {\bf 695} (2011) 317, [gr-qc/1008.2768].


\bibitem{avoid}
R.~Torres,
 Phys.\ Lett.\ B {\bf 733}, 21 (2014), [arXiv:1404.7655 [gr-qc]];
 R.~Torres and F.~Fayos,
 Phys.\ Lett.\ B {\bf 733}, 169 (2014), [arXiv:1405.7922 [gr-qc]].



\bibitem{bh2}
K.~Falls, D.~F.~Litim and A.~Raghuraman,
 Int.\ J.\ Mod.\ Phys.\ A {\bf 27}, 1250019 (2012), [arXiv:1002.0260 [hep-th]].


\bibitem{Koch:2013owa}
B.~Koch and F.~Saueressig,
 Class.\ Quant.\ Grav.\   {\bf 31} (2013) 015006, [hep-th/1306.1546];
B.~Koch, C.~Contreras, P.~Rioseco and F.~Saueressig,
 [hep-th/1311.1121].


\bibitem{Lauscher:2005qz}
O.~Lauscher and M.~Reuter,
 JHEP {\bf 0510} (2005) 050, [hep-th/0508202].


\bibitem{Reuter:2011ah}
M.~Reuter and F.~Saueressig,
JHEP {\bf 1112} (2011) 012, [hep-th/1110.5224].

\bibitem{Rechenberger:2012pm}
S.~Rechenberger and F.~Saueressig,
Phys.\ Rev.\ D {\bf 86} (2012) 024018, [hep-th/1206.0657].

\bibitem{Calcagni:2013vsa}
G.~Calcagni, A.~Eichhorn and F.~Saueressig,
Phys.\ Rev.\ D {\bf 87}, 124028 (2013), [hep-th/1304.7247].


\bibitem{Bonanno:2001xi}
A.~Bonanno and M.~Reuter,
Phys.\ Rev.\ D {\bf 65} (2002) 043508, [hep-th/0106133];
A.~Bonanno and M.~Reuter,
 Phys.\ Lett.\ B {\bf 527} (2002) 9, [astro-ph/0106468].


\bibitem{cosmofrank}
M.~Reuter and F.~Saueressig,
JCAP {\bf 0509}, 012 (2005), [hep-th/0507167].

\bibitem{elo}
E.~Bentivegna, A.~Bonanno and M.~Reuter,
 JCAP {\bf 0401}, 001 (2004), [astro-ph/0303150].


\bibitem{esposito}
A.~Bonanno, G.~Esposito, C.~Rubano and P.~Scudellaro,
Class.\ Quant.\ Grav.\  {\bf 24}, 1443 (2007), [gr-qc/0610012].


\bibitem{weinberginflation}
S.~Weinberg,
Phys.\ Rev.\ D {\bf 81}, 083535 (2010), [arXiv:0911.3165 [hep-th]].



\bibitem{c1}
  A.~Bonanno and M.~Reuter,
  Int.\ J.\ Mod.\ Phys.\ D {\bf 13}, 107 (2004)
  [astro-ph/0210472];
  M.~Reuter and H.~Weyer,
  JCAP {\bf 0412}, 001 (2004)
  [hep-th/0410119];
  M.~Reuter and H.~Weyer,
  Phys.\ Rev.\ D {\bf 70}, 124028 (2004)
  [hep-th/0410117].


\bibitem{c2}
  M.~Reuter and H.~Weyer,
  Int.\ J.\ Mod.\ Phys.\ D {\bf 15}, 2011 (2006)
  [hep-th/0702051];
  B.~F.~L.~Ward,
  Mod.\ Phys.\ Lett.\ A {\bf 23}, 3299 (2008)
  [arXiv:0808.3124 [gr-qc]];
  A.~Bonanno, A.~Contillo and R.~Percacci,
  Class.\ Quant.\ Grav.\  {\bf 28}, 145026 (2011)
  [arXiv:1006.0192 [gr-qc]].


\bibitem{c3}
  S.-H.~H.~Tye and J.~Xu,
  Phys.\ Rev.\ D {\bf 82}, 127302 (2010)
  [arXiv:1008.4787 [hep-th]];
  B.~F.~L.~Ward,
  PoS ICHEP {\bf 2010}, 477 (2010)
  [arXiv:1012.2680 [gr-qc]];
  R.~J.~Yang,
  Eur.\ Phys.\ J.\ C {\bf 72}, 1948 (2012)
  [arXiv:1108.0227 [gr-qc]];
  Y.~F.~Cai and D.~A.~Easson,
  Phys.\ Rev.\ D {\bf 84}, 103502 (2011)
  [arXiv:1107.5815 [hep-th]];
  A.~Contillo, M.~Hindmarsh and C.~Rahmede,
  Phys.\ Rev.\ D {\bf 85}, 043501 (2012)
  [arXiv:1108.0422 [gr-qc]].

\bibitem{Reuter:2012xf}
M.~Reuter and F.~Saueressig,
Lect.\ Notes Phys.\  {\bf 863}, 185 (2013), [arXiv:1205.5431 [hep-th]].


\bibitem{Bonanno:2008xp}
A.~Bonanno and M.~Reuter,
 J.\ Phys.\ Conf.\ Ser.\  {\bf 140}, 012008 (2008), [arXiv:0803.2546 [astro-ph]].


\bibitem{2007JCAP...08..024B}
A.~Bonanno and M.~Reuter,
 JCAP {\bf 0708}, 024 (2007), [arXiv:0706.0174 [hep-th]].

\bibitem{troy}
E.~P.~Tryon,
 Nature {\bf 246}, 396 (1973).



\bibitem{bianchi}
M.~Reuter and H.~Weyer,
 Phys.\ Rev.\ D {\bf 69}, 104022 (2004), [hep-th/0311196].


\bibitem{Kofinas:2015sna}
G.~Kofinas and V.~Zarikas,
 JCAP {\bf 1510}, no. 10, 069 (2015), [arXiv:1506.02965 [hep-th]].


\bibitem{Guberina:2002wt}%
B.~Guberina, R.~Horvat and H.~Stefancic,
 Phys.\ Rev.\ D {\bf 67}, 083001 (2003), [hep-ph/0211184].


\bibitem{Bauer:2005rpa}
F.~Bauer,
 Class.\ Quant.\ Grav.\  {\bf 22}, 3533 (2005), [gr-qc/0501078].


\bibitem{Frolov:2011ys}
A.~V.~Frolov and J.~Q.~Guo,
 [arXiv:1101.4995 [astro-ph.CO]].

\bibitem{Hindmarsh:2012rc}
M.~Hindmarsh and I.~D.~Saltas,
 Phys.\ Rev.\ D {\bf 86}, 064029 (2012), [arXiv:1203.3957 [gr-qc]].
%
\bibitem{Copeland:2013vva}
E.~J.~Copeland, C.~Rahmede and I.~D.~Saltas,
 Phys.\ Rev.\ D {\bf 91}, no. 10, 103530 (2015), [arXiv:1311.0881 [gr-qc]].

\bibitem{Manrique:2010am}
  E.~Manrique, M.~Reuter and F.~Saueressig,
  Annals Phys.\  {\bf 326}, 463 (2011)
  [arXiv:1006.0099 [hep-th]].

\bibitem{erdelyi}
A. Erdeleyi, {\it{Higher Transcendental Functions}}, Volume 1, McGraw-Hill Book Company, Inc., 1953.

\bibitem{abramowich}
M. Abramowitz and I.A. Stegun, {\it{Handbook of Mathematical Functions with Formulas, Graphs,
and Mathematical Tables}}, Dover Books on Mathematics, 1965.

\bibitem{Hindmarsh:2011hx}
  M.~Hindmarsh, D.~Litim and C.~Rahmede,
  JCAP {\bf 1107}, 019 (2011)
  [arXiv:1101.5401 [gr-qc]].

\bibitem{Ahn:2011qt}
  C.~Ahn, C.~Kim and E.~V.~Linder,
  Phys.\ Lett.\ B {\bf 704}, 10 (2011)
  [arXiv:1106.1435 [astro-ph.CO]].

\bibitem{Weinbergold}
S. Weinberg, {\textit{Gravitation and cosmology:
principles and applications of the general theory of relativity}}, New York: Wiley, 1972.

\bibitem{Sen:2000zk}
  S.~Sen and A.~A.~Sen,
  Phys.\ Rev.\ D {\bf 63}, 124006 (2001)
  [gr-qc/0010092].

\bibitem{zimdhalh}
  W.~Zimdahl, D.~J.~Schwarz, A.~B.~Balakin and D.~Pavon,
  Phys.\ Rev.\ D {\bf 64}, 063501 (2001)
  [astro-ph/0009353].

\bibitem{tetrdis}
  S.~Floerchinger, N.~Tetradis and U.~A.~Wiedemann,
  Phys.\ Rev.\ Lett.\  {\bf 114}, no. 9, 091301 (2015)
  [arXiv:1411.3280 [gr-qc]].

\bibitem{modphys}
  M.~S.~Berman and L.~A.~Trevisan,
  Int.\ J.\ Mod.\ Phys.\ D {\bf 19}, 1309 (2010)
  [gr-qc/0104060];
Berman, M.S. Marinho Jr, R.M. (2001), Astrophysics and Space Science, 278, 367.

\bibitem{Prigogine:1989zz}
  I.~Prigogine, J.~Geheniau, E.~Gunzig and P.~Nardone,
  Gen.\ Rel.\ Grav.\  {\bf 21}, 767 (1989).


\bibitem{freaza}
  M.~P.~Freaza, R.~S.~de Souza and I.~Waga,
  Phys.\ Rev.\ D {\bf 66}, 103502 (2002).

\bibitem{waga}
  R.~O.~Ramos, M.~V.~d.~Santos and I.~Waga,
  Phys.\ Rev.\ D {\bf 89}, no. 8, 083524 (2014)
  [arXiv:1404.2604 [astro-ph.CO]].

\bibitem{Israel:1979wp}
  W.~Israel and J.~M.~Stewart,
  Annals Phys.\  {\bf 118}, 341 (1979).


\bibitem{Aliferis:2014ofa}
  G.~Aliferis, G.~Kofinas and V.~Zarikas,
  Phys.\ Rev.\ D {\bf 91} (2015) 4, 045002, hep-ph/1406.6215.


\bibitem{Kofinas:2011pq}
  G.~Kofinas and V.~Zarikas,
  Eur.\ Phys.\ J.\ C {\bf 73} (2013) 4, 2379, hep-th/1107.2602.

\bibitem{Braun:2010tt}
  J.~Braun, H.~Gies and D.~D.~Scherer,
  Phys.\ Rev.\ D {\bf 83}, 085012 (2011)
  [arXiv:1011.1456 [hep-th]].



\end{thebibliography}
\end{document}